\documentclass[12pt]{article}
\usepackage{amssymb,amsmath,amsfonts}
\usepackage{graphicx}
\graphicspath{{Figures/}}
\usepackage{color}
\numberwithin{equation}{section}

\newcommand{\be}{\begin{equation}}
\newcommand{\ee}{\end{equation}}
\newcommand{\bea}{\begin{eqnarray}}
\newcommand{\eea}{\end{eqnarray}}

\newcommand{\e}{{\rm e}}

\renewcommand{\d}{{\rm d}}
\renewcommand{\i}{{\rm i}}

\newcommand{\grintl}{[\kern-.18em [}
\newcommand{\grintr}{]\kern-.18em ]}

\newcounter{resultcounter}[section]

\newtheorem{thm}[resultcounter]{Theorem}
\newtheorem{lem}[resultcounter]{Lemma}
\newtheorem{prop}[resultcounter]{Proposition}
\newtheorem{cor}[resultcounter]{Corollary}
\newtheorem{definition}[resultcounter]{Definition}

\def\bed{\begin{definition}}
	\def\eed{\end{definition}}

\newcommand{\scalprod}[2]{\left\langle {#1}, {#2}\right\rangle}
\newcommand{\bbbone}{\mathchoice {\rm 1\mskip-4mu l} {\rm 1\mskip-4mu l}
	{\rm 1\mskip-4.5mu l} {\rm 1\mskip-5mu l}}

\newcommand{\s}{{\rm S}}
\renewcommand{\r}{{\rm R}}

\begin{document}

		\title{\vspace*{-1cm} Production of Entanglement Entropy 
			by Decoherence}
			

		\author{ M. Merkli\footnote{Department of Mathematics and Statistics, Memorial University of Newfoundland, St. John's, NL, Canada A1C 5S7; merkli@mun.ca}\and G.P. Berman\footnote{Theoretical Division T-4, B-11, Los Alamos National Laboratory and the New Mexico Consortium,  Los Alamos, NM, 87544, USA; gpb@lanl.gov}  \and R.T. Sayre\footnote{Bioscience Division, B-11, Los Alamos National Laboratory and the New Mexico Consortium, 100 Entrada Dr., Los Alamos, NM 87544, USA; rsayre@newmexicoconsortium.org} \and X. Wang\footnote{Quantum Strategics, San Diego, California, USA; xidiwang2003@yahoo.com} \and  A.I. Nesterov\footnote{Departamento de Fisica, CUCEI, Universidad de Guadalajara, Av. Revoluci\'on 1500, Guadalajara, CP 44420, Jalisco, M\'exico; nesterov@cencar.udg.mx}}

		\maketitle
		\small LA-UR-17-20823
	
\begin{abstract}
We examine the dynamics of entanglement entropy of all parts in an open system consisting of a two-level dimer interacting with an environment of oscillators. The dimer-environment interaction is almost energy conserving. We find the precise link between decoherence and production of entanglement entropy.  We show that not all environment oscillators carry significant entanglement entropy and we identify the oscillator frequency regions which contribute to the production of entanglement entropy. Our results hold for arbitrary strengths of the  dimer-environment interaction, and they are mathematically rigorous. 
\end{abstract}
			
\section{Introduction}

Quantum systems in contact with an environment (reservoir) typically undergo decoherence. An initial pure system state becomes mixed and hence `more classical' due to the system-reservoir interaction \cite{JoosBook}. This loss of purity (or, `quantumness') is mediated by entanglement between the system and its environment and can be quantified by the notion of entanglement entropy (EE). First introduced in \cite{Bennett}, the examination of EE is receiving enormous attention from researches in different fields of science, ranging from solid state physics and the theory of quantum computation and information to quantum field-  and black hole theory  \cite{Cardy1,Cardy2,Bai,Thermal,Raamsdonk,Cardy3,JM,PeschelReview,Tasaki,M1,M2} (see also references therein).

The goal of the present work is to study the time evolution of EE in a simple open system, a two-level quantum system interacting with a field of quantum oscillators. We adopt two premises:
\begin{itemize}
\item[(1)] The initial state of the combined total system plus environment is a {\em pure} state.

\item[(2)] The interaction between the system and the environment is (almost) {\em energy conserving}, meaning that the system Hamiltonian is (almost) a constant of motion. 
\end{itemize}

Assumption (1) allows us to study quantum correlations between different parts of the system-environment complex using entanglement entropy (see the paragraph below). In the general theory of open quantum systems, reservoirs are often postulated to be {\em  initially in thermal equilibrium states}, which are mixed states and which do not fall into the category of assumption (1). However, in  our view, it is not necessarily clear why one should {\em start off} with initial states in thermal equilibrium, as indeed, equilibrium can be shown to {\em emerge} by reducing a {\em pure state} to a subsystem. In view of this, premise (1) is  rather natural. We comment on this phenomenon in Section \ref{ultimsect}.

It is well known that decoherence happens more quickly, sometimes much more quickly, than relaxation (dynamics of populations) in many open systems \cite{BP, JoosBook, AlickiLendi}. Since relaxation is produced by energy-exchange processes, but decoherence can be brought about by energy-conserving ones, our assumption (2) is reasonable for time scales not exceeding the system decoherence time. It follows that our approach allows a good description of the {\em relation between entanglement production and decoherence}, which is the main topic of the present work.

Before  giving an outline of our main results in Section \ref{subs11}, we explain the notion of EE.

\medskip

{\bf Entanglement entropy (EE).\ } A quantum state (density matrix) $\rho$ is called a {\em pure state} if it has rank one, or equivalently, if $\rho = |\psi\rangle\langle\psi|$ for some vector $\psi$. A state which is not pure is called a {\em mixed state}. The {\em von Neumann entropy} of a state $\rho$ is defined by  $S(\rho) = -{\rm Tr}(\rho\ln\rho)$, which is a non-negative number. The equivalence  

\medskip

\centerline{$\rho$ is pure  $\Longleftrightarrow$ $S(\rho)=0$ }

\medskip
\noindent
makes the von Neumann entropy a good measure for the purity of a state. Given a bipartite quantum system $A+B$, a pure state determined by a vector $\psi$ is called {\em entangled} if it is not possible to find vectors $\psi_A$ and $\psi_B$ of the individual subsystems, such that $\psi = \psi_A\otimes\psi_B$. Equivalently, $\psi$ is called {\em disentangled} (or not entangled), or a {\em product state}, if $\psi=\psi_A\otimes\psi_B$ for some subsystem vectors $\psi_A$, $\psi_B$. 


 The following is a basic relation between purity and entanglement. Let $\psi$ be a pure state of a bipartite quantum system $A+B$. The reduced states (density matrices) of subsystems $A$ and $B$ are defined by $\rho_A={\rm Tr}_B |\psi\rangle\langle\psi|$ and $\rho_B={\rm Tr}_A |\psi\rangle\langle\psi|$ (partial traces). Then the following equivalence holds,

\medskip

\centerline{$\rho_A$ is  pure  $\Longleftrightarrow$ $\psi$ is not entangled.}

\medskip
\noindent
This is also equivalent to $\rho_B$ being a pure state. It is important that the state of the  system $A+B$ be {\em pure} for the above equivalence to hold; indeed if $\rho=\sigma_A\otimes\sigma_B$ is the product of two mixed states, then the reduced state of $A$,  $\rho_A=\sigma_A$, is not pure. Combining the above two equivalences leads to the following result. 

\medskip

\centerline{\em Let $\psi$ be a pure state of a bipartite quantum system $A+B$ and let $\rho_A={\rm Tr}_B|\psi\rangle\langle\psi|$. Then}

\centerline{\em $S(\rho_A)=0$  $\Longleftrightarrow$ $\psi$ is not entangled.}

\medskip
\noindent
This  motivates the {\bf definition of entanglement entropy} 
\cite{Bennett,Peschel} to be $S(\rho_A)$, the von Neumann entropy of the reduction of the pure state of a bipartite system to one of its parts. As it turns out, when $\rho_A$ and $\rho_B$ are the reductions to the subsystems $A$ and $B$, obtained from a pure state of $A+B$, then $S(\rho_A)=S(\rho_B)$. So it does not matter of which subsystem state we take the von Neumann entropy, in the definition of entanglement entropy.


\subsection{Outline of main results}
\label{subs11}

In this section, we present the initial state and the Hamiltonians of our models \eqref{22}-\eqref{0032} and we define the reduced density matrices for the dimer \eqref{strace} and (all parts of) the reservoir \eqref{resredstat}. We then present concrete expressions for the dimer decoherence in subsection \ref{dimdecsect} and for the dynamics of the entanglement entropy in subsection \ref{EEEdynsect}. In the latter, we show that an increase in the dimer EE is equivalent to an increase in the dimer decoherence. In subsection \ref{applisect}, we mention applications to a noisy qubit and to the photosynthetic light harvesting complex.

\medskip

We consider a dimer (a two-level system) in contact with $N$ oscillators. The initial state of the entire system is 
\begin{equation}
\label{22}
|\Psi_0\rangle = |\psi_\s\rangle\otimes |\alpha_1\rangle \otimes|\alpha_2\rangle \cdots\otimes |\alpha_N\rangle,
\end{equation}
where $|\psi_\s\rangle = a|\!\uparrow\rangle +b |\!\downarrow\rangle$ (with $|a|^2=p$, $|b|^2=1-p$) is an arbitrary pure state of the dimer  and
$$
|\alpha_j\rangle = \e^{-|\alpha_j|^2/2} \sum_{n\ge 0} \frac{\alpha_j^n}{\sqrt{n!}} |n\rangle
$$
is a single-oscillator coherent state, $\alpha_j\in\mathbb C$. While we discuss here an environment of oscillators initially in coherent states, many of our formulas are readily expressed for general, pure or mixed initial states for the oscillators (see Section \ref{FinModeResSect}). We analyze two different models. In the first one, called the {\bf $X$-interaction} model, the total Hamiltonian is given by
\begin{equation}
\label{001}
H_{X} = \tfrac12 \Omega\sigma_z + \sum_{j=1}^N \omega_j a^\dagger_ja_j +\lambda\sigma_z\otimes \sum_{j=1}^N g_j (a^\dagger_j+a_j),
\end{equation}
where $\Omega,\omega_j >0$, $\lambda\in\mathbb R$ and $\sigma_z$ is the Pauli $z$-matrix. The operators $a^\dagger_j$ and $a_j$ are the creation and annihilation operators associated with the oscillator labeled by $j$, satisfying the canonical commutation relations $a_j a^\dagger_k-a^\dagger_k a_j=\delta_{kj}$ (Kronecker delta). The collection of complex numbers  $\{g_j\}_{j=1}^N$ is called the {\em form factor}. The size of $|g_j|$ measures how strongly the oscillator labeled by $j$ is coupled to the dimer.\footnote{To our knowledge, a mathematically controllable theory for the full system-reservoir dynamics in the presence of an additional term $V\sigma_x$ in the Hamiltonian \eqref{001} is not available yet. Nevertheless, the exact non-Markovian master equation for the dimer alone, evolving according to \eqref{001}  with $V\sigma_x$ added, and where initially the dimer and reservoir are disentangled and the latter is  in {\em thermal equilibrium}, has been derived recently in \cite{Ferialdi}. 
} The second model we discuss is described by a Hamiltonian with a density-, or {\bf $D$-interaction}, given by 
\begin{equation}
\label{0032}
H_D = \tfrac12 \Omega\sigma_z +\tfrac 12 V\sigma_x +\sum_{j=1}^N\omega_j a^\dagger_ja_j +\lambda\sigma_z\otimes\sum_{j=1}^N g_j a^\dagger_ja_j +\mu\sigma_x\otimes\sum_{j=1}^N f_j a^\dagger_ja_j.
\end{equation}
Here, $\sigma_x$ is the Pauli $x$-matrix and we assume in this model that $\Omega, \omega_j, \lambda,\mu,g_j,f_j > 0$. 

Note that the dimer energy, $\tfrac12 \Omega\sigma_z$, is a conserved quantity for the dynamics generated by $H_X$ as well as for that generated by $H_D$ with $V=\mu=0$. Models in which the dimer energy is conserved are called {\em energy conserving}. The oscillator energy, $\sum_{j=1}^N \omega_j a^\dagger_ja_j$, is also conserved for $H_D$, but not for $H_X$.  

We denote by $\rho_\s(t)$ be the {\bf reduced dimer density matrix}, obtained by tracing out all oscillator degrees of freedom. It is given by
\begin{equation}
\label{strace}
\rho_\s(t) ={\rm Tr}_{\rm oscillators} \ \e^{-\i tH} \, |\Psi_0\rangle\langle \Psi_0|\,  \e^{\i tH},
\end{equation}
where $\Psi_0$ is the initial pure dimer-reservoir state \eqref{22}. 
The matrix elements are denoted by $[\rho_\s(t)]_{ij}$ ($i=1$ being associated with the eigenvector $|\!\uparrow\rangle$ of $\sigma_z$ with eigenvalue $+1$). The reduced oscillator density matrix is defined as follows. Let $J\subseteq \mathbb R$ be a window of frequencies of interest. For a given distribution of the oscillator frequencies $\{\omega_1,\ldots,\omega_N\}$, we have, say, $K$ discrete frequencies lying inside $J$. Denote by $\rho_J(t)$ the {\bf reduced oscillator density matrix} of all oscillators with frequencies inside $J$. Similarly to \eqref{strace}, $\rho_J(t)$ is obtained from the full density matrix by tracing out the degrees of freedom of the dimer and of all oscillators with (the $N-K$ discrete) frequencies lying outside $J$,
\begin{equation}
	\label{resredstat}
\rho_J(t) = {\rm Tr}_{{\rm system\, + \, oscillators \ \ell \ with\  \omega_\ell\not\in} J} \ \ \e^{-\i tH} \, |\Psi_0\rangle\langle \Psi_0|\,  \e^{\i tH}.
\end{equation}
The sets of non-vanishing eigenvalues of $\rho_\s(t)$ and of  $\rho_J(t)$ are called the {\bf entanglement spectra} of the respective reduce density matrices \cite{LiHaldane}.

\subsubsection{Dimer decoherence}
\label{dimdecsect}

For finitely many oscillators (frequencies), the dynamics of the dimer and of the oscillators is quasi-periodic in time. As the oscillator frequency spectrum becomes denser, recurrence times are longer and in the (idealized) limit of a {\em continuum of oscillator frequencies} (see Section \ref{contmodesect}), one observes irreversible dynamics. Since the system energy is conserved for the Hamiltonians \eqref{001} and \eqref{0032} with $V=\mu=0$ (the Hamiltonian $\tfrac12 \Omega \sigma_z$ commutes with the total Hamiltonian), the populations of the system -- the diagonal matrix elements in the energy basis --  are constant in time. In Theorems \ref{lem04} and \ref{lem4'} we analyze the dimer energy conserving dynamics of $\rho_\s(t)$ and $\rho_J(t)$, for any frequency window $J$. We show that
\begin{itemize}
\item The oscillator density matrix $\rho_J(t)$ has rank two ($t>0$). Its entanglement spectrum consists of two eigenvalues which we find explicitly for any $t> 0$ and any $N$.
\item The  off-diagonal dimer density matrix element evolves according to the law
\begin{equation}
\label{caldo}
\big| \,[\rho_{\rm S}(t)]_{12}\, \big|  =  |s_{[0,\infty)}(t)| \ \big| \, [\rho_{\rm S}(0)]_{12}\, \big|\, ,
\end{equation}
where, for any $J\subseteq [0,\infty)$,
\begin{equation}
\label{thesj}
s_J(t) =
\left\{
\begin{array}{ll}
\e^{-4\lambda^2\int_J h_X(\omega) \ \, \frac{1-\cos(\omega t)}{\omega^2} \, d\omega}, & \mbox{$X$-interaction model}\\
\e^{-\int_J h_D(\omega) \, [1-\e^{2\i t\lambda g(\omega)}]\, d\omega}, &  \mbox{$D$-interaction model}
\end{array}
\right.
\end{equation}
Here,
\begin{equation}
\label{theh}
h_X(\omega) = \tfrac{dn}{d\omega} \, |g(\omega)|^2, \qquad\mbox{and} \qquad h_D(\omega) = \tfrac{dn}{d\omega} \,  |\alpha(\omega)|^2
\end{equation}
and $\tfrac{dn}{d\omega}(\omega)$ is the frequency density of the reservoir initial state at the frequency $\omega$. In the continuous mode limit, the decoherence factor $0\le |s_{[0,\infty)}(t)| \le 1$ satisfies (see Propositions \ref{lem5} and \ref{prop2})
\begin{equation}
\label{cald}
|s_{[0,\infty)}(t)|=
\left\{
\begin{array}{ll}
\e^{-2\pi t  \lambda^2 h_X(0)\, [1+o_t]}, & \mbox{$X$-interaction model}\\
\e^{-\int_0^\infty h_D(\omega) d\omega}\ \big[\, 1+ O\big( (t_{\rm pd}/t)^2\big)\, \big], & \mbox{$D$-interaction model}
\end{array}
\right.
\end{equation}
and $o_t$ is a quantity converging to zero as $t\rightarrow\infty$. Relations \eqref{caldo} and \eqref{cald} show that decoherence is {\em full} for the $X$-interaction (provided $h_X(0)\neq 0$), meaning that $[\rho_\s(t)]_{12}\rightarrow 0$ as $t\rightarrow\infty$. For the $D$-interaction, decoherence is only partial and happens on a partial decoherence time scale $t_{\rm pd}$, explicitly given in \eqref{57} below.

\item 
While \eqref{caldo} and \eqref{cald} hold for dimer energy conserving models, we examine in Theorems \ref{thm01} and \ref{thm1a} the evolution generated by \eqref{0032} when additionally, $V/\Omega$ takes  small non-vanishing values. We isolate the main term in the dynamics as well as the first order correction $\propto V/\Omega$, and we control the error on time scales up to $t_{\rm pd}$. We show that the energy-exchange interaction produces a correction to the (strictly positive) entanglement entropy caused by the energy conserving interaction, whose sign fluctuates in time and depends on the initial dimer state (see Section \ref{almostsect})

\end{itemize}

\subsubsection{Dynamics of the entanglement entropy}
\label{EEEdynsect}

We show in Theorems \ref{lem04} and \ref{lem4'} that the (nonzero) spectrum of the reduced density matrices $\rho_\s(t)$ and $\rho_J(t)$, in the energy-conserving situations, are given by
\begin{eqnarray}
\label{ECspectra}
{\rm spec}\big( \rho_J(t)\big) &=& \tfrac12 \pm \tfrac12 r_{J}(t)\\
{\rm spec}\big( \rho_\s(t)\big) &=& {\rm spec}\big( \rho_{J=[0,\infty)}(t)\big),
\label{EC'}
\end{eqnarray}
where
\begin{equation}
\label{rj}
r_J(t) = \sqrt{1 -4p(1-p)(1-|s_J(t)|^2)},
\end{equation}
with $p=|\langle \uparrow\!|\psi_\s\rangle|^2$. Note that \eqref{EC'} follows immediately from \eqref{ECspectra}. Indeed, by partitioning the whole system into parts $A$=dimer and $B$=oscillators, the entanglement entropy of $A$ automatically equals that of $B$ (see the introduction). The entanglement entropy is thus
\begin{equation}
\label{SJT}
S_J(t)\equiv S\big(\rho_J(t)\big) = -\big(\tfrac12 +\tfrac12 r_J(t) \big)\ln\big(\tfrac12 +\tfrac12 r_J(t)\big) -\big(\tfrac12 -\tfrac12 r_J(t) \big) \ln\big(\tfrac12 -\tfrac12 r_J(t) \big).
\end{equation}
This gives us the following results.
\begin{itemize}
\item[1.] {\bf The entanglement entropy $S_J$ is a monotonically decreasing function of the decoherence factor $|s_J|$.} The parameter $|s_J|$ (see \eqref{thesj}) takes values in the closed interval $[0,1]$. The entanglement entropy $S_J$ is monotonically decreasing in the difference of the eigenvalues, $|r_J|$, hence by \eqref{rj}, $S_J$ is a monotonically decreasing function of $|s_J|$ (meaning that $\frac{d}{d|s_J|}S_J(|s_J|)<0$). The value of $S_J$ is  maximal ($S_J=\ln 2$) when both eigenvalues are equal, for $r_J=0$. It is minimal ($S_J=0$) if $r_J=1$. In particular, if $p=0$ or $p=1$, then $r_J(t)=1$ for all $J$ and all $t$ and no entanglement is ever created. We have $s_J(0)=1=r_J(0)$ and $S_J(0)=0$, reflecting the fact that the initial state is  pure and entirely disentangled.

\item[2.] {\bf Dimer decoherence and dimer EE production.} According to \eqref{caldo} and \eqref{cald}, we say the dimer decoherence is large provided $|s_{[0,\infty)}(t)|$ (off-diagonal) is small. This shows that an increase in the dimer entanglement entropy is necessarily accompanied by an increase in the dimer decoherence.  We say the dimer undergoes {\em full decoherence} if $\lim_{t\rightarrow\infty} |s_{[0,\infty)}(t)| =0$ and {\em partial decoherence}  if $\lim_{t\rightarrow\infty} |s_{[0,\infty)}(t)| > 0$.
If the dimer undergoes full decoherence, then its final EE, $S_{[0,\infty)}(\infty)$, is given by \eqref{SJT} with $r_J=r(p) = \sqrt{1-4p(1-p)}$. As a function of $p\in[0,1]$, $S_{[0,\infty)}(\infty)$ is symmetric around the point $p=1/2$ and increases monotonically from the value $0$ for $p=0$ to the value $\ln 2$ for $p=1/2$. Full decoherence happens in the energy conserving $X$-interaction system (Proposition \ref{lem5}), while for the energy-conserving $D$-interaction, we have partial decoherence (Proposition \ref{prop2}).

\item[3.] {\bf Localization of EE in the oscillators. } Consider first the $X$-interaction model. We show in Proposition \ref{lem5} (B) below that the value of $|s_J(t)|$ for $J=[0,\omega_1]$ with arbitrary $\omega_1>0$, and hence the entanglement entropy $S(\rho_J(t))$, are determined for large times entirely by the coupling function at the frequency zero, $h_X(0)$ (see \eqref{theh}). In Proposition \ref{lem5} (C) we show that as the interval $J$ moves away from the low frequency oscillator region, the entanglement entropy of $\rho_J$ decreases. More precisely, let $J=[\omega_0,\omega_1]$ with $0<\omega_0<\omega_1$. When $\omega_0$ in increases, the quantity $|s_J(t)|$ increases as well and hence the entanglement entropy of $\rho_J(t)$ decreases. This shows that {\bf for the $X$-interaction, the entanglement entropy in the oscillators is concentrated in the low-frequency region}. For the $D$-interaction, a different picture emerges from our Proposition \ref{prop2} below.  By point 1. above, $S_J$ decreases monotonically in $s_J$, which by \eqref{thesj} itself is a decreasing function in the size of the considered frequency window. Assuming for the sake of the argument that $h_D(\omega) \approx h_{D,0}$ is roughly constant over a region $\omega\in [0,\omega_c]$, we see from \eqref{cald} that for large times, we have $s_J \approx \e^{-|J| h_{D,0}}$, where $|J|$ is the length of the interval  $J\subset [0,\omega_c]$. It follows that the oscillator entanglement entropy $S(\rho_J^0)$ is translation invariant for large times. Namely, {\bf for the $D$-interaction, the entanglement entropy in the oscillators with frequencies in $J$ only depends on the size of $J$, not on the location of $J$}. This suggests that in the $D$-interaction model, {\em all oscillators contribute to the dynamical process}.
\end{itemize}

\subsubsection{Applications}

{\bf (A) The decohering qubit}

In quantum computation and information theory, the dimer represents a qubit and our energy conserving interaction is the coupling to a purely dephasing noise. There is a huge literature on the dynamics on qubits in quantum computation, we mention only the paper \cite{Palma} and the books \cite{BP, JoosBook, Schlosshauer}. Characteristic times scales for the dephasing of superconducting qubits in laboratories have been pushed to $\tau_D\sim 20\mu$s, while the relaxation time scale  is $\tau_R\sim 60$$\mu$s, \cite{Steffen}. The dynamics can be viewed as purely dephasing, driven by an energy conserving Hamiltonian, for a long time $t\lesssim \tau_D$. 

\medskip

\noindent
{\bf (B) The photosynthetic light harvesting complex}
\label{applisect}

In light harvesting antenna complexes of photosynthetic organisms, dimers based on two chlorophyll molecules,  Chl$b$ (donor) and Chl$a$ (acceptor), have been proposed to play a critical role in regulating energy (exciton) transfer and dissipation \cite{Lloyd,Scholes,Chin,Engel,Fleming1,Fleming2}. The chlorophyll molecules are embedded in a protein-solvent environment and form an open quantum system. The chlorophyll dimer is described by a two-level system representing the two excited energy levels of the chlorophyll molecules. The ground states of the chlorophylls are not included in the description, as it is assumed that the time of the excitation transfer between the two excited levels is much shorter than either the time of fluorescence or the time of decay of the exciton to the environment \cite{Engel}. When a chlorophyll  dimer is populated by light absorption, it undergoes the processes of decoherence and relaxation. Decoherence happens much faster than relaxation (evolution of the populations, i.e.,  diagonal density matrix elements), and so we are in the energy conserving regime for the dynamics and our theory  applies. Typical decoherence time scales are  $\tau_D\sim~300$fs while relaxation times are  $\tau_R\sim  10$ps.


Current technologies allow detailed observations and measurements of decoherence effects in photosynthetic light harvesting complexes \cite{Engel, Fleming2}. Therefore, a measurement of entanglement entropy in the dimer is experimentally accessible. Since the dimer entanglement entropy equals the entanglement entropy of the environment (see the introduction above), one has experimental access to characteristics of the environment as well.

\subsection{Continuous mode limit}
\label{contmodesect}

The dynamics generated by either of $H_X$ or $H_D$ is quasi-periodic, and stays so even when $N\rightarrow\infty$, if the oscillator frequencies $\omega_j$ form a discrete set. Irreversible dynamics (of the dimer and of the oscillators) emerges only in the presence of a {\em continuum} of oscillator frequencies. We will refer to the procedure of taking the oscillator frequencies closer and closer together as the {\em continuous mode limit}.\footnote{This does not mean that we create `modes' in the sense of delocalized standing waves in the reservoir. Rather, it means that we increase the number of atoms at varying frequencies making up the environment, thus getting a very dense frequency spectrum.} The continuous mode limit is taken as follows. Let  $f_j=f(\omega_j)$ be a function of the frequency $\omega_j$. Then
\begin{equation}
\label{25}
\sum_{j=1}^N f(\omega_j) = \sum_{j=1}^{N'} n(\omega_j) f(\omega_j) = \sum_{j=1}^{N'} \frac{n(\omega_j)}{\Delta\omega_j} f(\omega_j)\Delta\omega_j \ \longrightarrow\ \int_0^\infty \frac{dn(\omega)}{d\omega} f(\omega)d\omega,
\end{equation}
where $N'$ is the number of distinct, discrete frequencies and  $n(\omega_j)$ is the number of oscillators having the same frequency $\omega_j$, $\Delta\omega_j =\omega_{j+1}-\omega_j$ and $\tfrac{dn(\omega)}{d\omega}$ is the density of oscillators at the given continuous frequency $\omega$. In what follows, when we refer to {\em functions} of the continuous frequency, $\alpha(\omega)$, $f(\omega)$, $g(\omega)$, we assume that the continuous mode limit has been performed.

The procedure \eqref{25} tells us how to obtain the continuous mode limit of suitable {\em sums} of a frequency dependent function. However, not all quantities of interest are of this form! For example, the reduced density matrix $\rho_J(t)$ of all oscillators with frequencies lying inside $J$ is an operator acting on the Hilbert space $\otimes_{j=1}^K {\mathcal H}_{\rm osc}$, where ${\mathcal H}_{\rm osc}$ is the pure state space of a single oscillator and $K$ counts the number of elements in $J$. We derive in Lemma \ref{lem2} an explicit form of $\rho_J(t)$. It cannot be expressed merely in terms of quantities of the form \eqref{25}. What then is the continuous mode limit of the state $\rho_J(t)$? This is not an easy question. It is not even clear on {\em what Hilbert space} this continuous mode state should act. Indeed, as we make the oscillator frequency spectrum denser and denser, the window $J$ will contain more and more of the discrete frequencies. In the wanted limit, the cardinality $K$ of the number of modes inside $J$ will increase to  ${\mathfrak c} =|\mathbb R|$, the cardinality of $\mathbb R$. Then we do not even have a candidate for a Hilbert space, since we do not know how to make sense of $\otimes_{j=1}^K {\mathcal H}_{\rm osc}$ in this limit! (How should we take a tensor product of uncountably many Hilbert spaces?) Nevertheless, there is a method of constructing the state and its Hilbert space in the continuous mode limit. It is based on the {\em Gelfand-Naimark-Segal} construction for linear functionals on $C^*$-algebras (i.e., states). This construction has been carried out in \cite{JM} for a coherent state reservoir. However, in the present context, as we will show below, the {\em  entanglement spectrum} of $\rho_J(t)$ can be expressed purely in terms of quantities to which we can apply the procedure \eqref{25}. Therefore, the mathematically more involved formalism of \cite{JM} is not needed for the considerations of the present work.

\section{Dynamics of subsystems for finite $N$}
\label{sect2}

\subsection{Energy conserving interactions}
\label{conssect}

{\bf Reduction of the rank of the oscillator density matrix.\ }
The structure of the reduced density matrices is easy to understand in general for the energy conserving situation. Suppose a $d$-dimensional system is coupled to a reservoir, described by an interacting Hamiltonian $H=H_\s + H_\r+ H_{\s\r}$, where $H_{\s\r}$ commutes with $H_\s$. Let $|\psi\rangle = |\psi_\s\rangle\otimes|\psi_\r\rangle$ be a pure system-reservoir initial state. Expanding $|\psi_\s\rangle = \sum_{k=1}^d c_k |\phi_k\rangle$, where  $\{ |\phi_k\rangle\}_{k=1}^d$ is an orthonormal basis of eigenvectors of $H_\s$,  we get 
\begin{equation}
\label{simple1}
|\psi(t)\rangle = \e^{-\i tH}|\psi\rangle = \sum_{k=1}^dc_k  |\phi_k\rangle\otimes \e^{-\i tH_k}|\psi_\r\rangle, 
\end{equation}
with $H_k$ a Hamiltonian acting on the reservoir alone. Indeed, because $H_\s$ and $H_{\s\r}$ commute, the eigenvectors $|\phi_k\rangle$ of $H_\s$ also diagonalize the system part of the operator $H_{\s\r}$, meaning that $H_{\s\r}(|\phi_k\rangle\otimes|\psi_\r\rangle) = |\phi_k\rangle\otimes (H'_{\s\r, k}|\psi_\r\rangle)$ for any reservoir state $|\psi_\r\rangle$, and where $H'_{\s\r, k}$ is an operator acting purely on the reservoir Hilbert space.\footnote{\label{footnote1} If $H_{\s\r} =  G\otimes\Phi$ with $G$ and $\Phi$ acting on the system and on the reservoir, respectively, then $H_{\s\r, k}'=g_k\Phi$, where $G|\phi_k\rangle = g_k|\phi_k\rangle$.}  Then $H_k = E_k +H_\r+H_{\s\r, k}'$, where $E_k$ is the energy of $|\phi_k\rangle$ (i.e., $H_\s|\phi_k\rangle = E_k |\phi_k\rangle$). 

It follows from \eqref{simple1} that the reduced reservoir density matrix has the form
\begin{equation}
\label{simple2}
\rho_\r(t) = {\rm Tr}_\s |\psi(t)\rangle\langle\psi(t)| = \sum_{k=1}^d |c_k|^2 \e^{-\i tH_k} |\psi_\r\rangle\langle\psi_\r|  \e^{\i tH_k}.
\end{equation}
Even though $\rho_\r(t)$ is a density matrix acting on an  (in general) infinite-dimensional reservoir Hilbert space, \eqref{simple2} shows that its {\em rank} is at most $d$, the dimension of the small system. Since the initial state is pure, this fact is also immediately implied by the Schmidt decomposition theorem, which shows that the ranks of $\rho_\s(t)$ and $\rho_\r(t)$ are equal, where $\rho_\s(t)$ is the reduced system density matrix. Assume now that the reservoir consists of $N$ uncoupled subsystems (e.g., oscillators), so that $H_\r = \sum_{j=1}^N H_{\r,j}$, and that the coupling to the system is of the form $\sum_{j=1}^N G_j\otimes\Phi_j$, where $G_j$ and $\Phi_j$ are coupling operators acting on the system and on the $j$th reservoir subsystem (oscillator), respectively. Then the operator $H_k$ in \eqref{simple1} has the form $H_k = \sum_{j=1}^N H_{k,j}$, where $H_{k,j}$ acts on the $j$th reservoir subsystem only (see also footnote \ref{footnote1}). For unentangled initial states $|\psi_\r\rangle = |\psi_{\r,1}\rangle \otimes\cdots\otimes |\psi_{\r,N}\rangle$ expression \eqref{simple2} becomes
\begin{equation}
\label{simple3}
\rho_\r(t) = \sum_{k=1}^d |c_k|^2 \textstyle \bigotimes_{j=1}^N \e^{-\i tH_{k,j}} |\psi_{\r,j}\rangle\langle \psi_{\r,j}|\e^{\i tH_{k,j}}.
\end{equation}
Therefore, reducing $\rho_\r(t)$ to a collection $J$ of reservoir subsystems (oscillators) results in the density matrix
\begin{equation}
\label{simple4}
\rho_{J}(t) = {\rm Tr}_{\{1,\ldots,N\}\backslash J} \ \rho_\r(t) = \sum_{k=1}^d |c_k|^2 \textstyle \bigotimes_{j\in J} \e^{-\i tH_{k,j}} |\psi_{\r,j}\rangle\langle \psi_{\r,j}|\e^{\i tH_{k,j}}.
\end{equation}
This shows that the reduced density matrix of any part of the reservoir has rank at most $d$ as well. For general systems with energy exchange, starting with a pure initial state, it is still true that whole reservoir density matrix $\rho_\r(t)$ has rank $\le d$ (again, just by the Schmidt decomposition theorem). However, the rank of $\rho_J(t)$ might {\em exceed} $d$. Indeed, $\rho_J(t)$ is obtained from $\rho_R(t)$ by tracing out parts of the reservoir degrees of freedom, an operation which may increase the rank (for instance, the partial trace of a pure state is generally a mixed state, hence taking the partial trace generally increases the rank). This can also be understood from a different point of view: switching on an energy exchange term (even a small one) shifts the spectrum of $\rho_J(t)$ and may cause some of its zero eigenvalues to become nonzero, hence increasing the rank of $\rho_J(t)$.

\medskip

{\bf Energy conserving dynamics.\ }
We examine the reduced dynamics of the dimer and (parts of) the reservoir under the dynamics generated by \eqref{001} and \eqref{0032} for $V=\mu=0$. Given $J\subseteq [0,\infty)$, we define the quantity (the discrete frequency analogue of \eqref{thesj})
\begin{equation}
\label{thesj.1}
s_J(t) =
\left\{
\begin{array}{ll}
	\exp \Big[ -4\lambda^2\sum_{\{j\, :\, \omega_j\in J\}} |g_j|^2  \, \frac{1-\cos(\omega_j t)}{\omega_j^2} \Big],  & \mbox{$X$-interaction model}\\
 & \\
	\exp\Big[ -\sum_{\{j\, :\, \omega_j\in J\}} |\alpha_j|^2 (1-\e^{2\i t\lambda g_j}) \Big] , &  \mbox{$D$-interaction model}
\end{array}
\right.
\end{equation}
We  prove in Theorems \ref{lem04}, \ref{lem4'} below that the reduced reservoir density matrix of all oscillators with frequencies inside $J$ is a rank-two operator, described in a suitable  basis by the matrix
\begin{equation}
\label{17.1}
\rho_J(t) = 
\begin{pmatrix}
p+(1-p)|s_J(t)|^2 & (1-p) s_J\sqrt{1-|s_J(t)|^2}\\
(1-p)\bar s_J(t)\sqrt{1-|s_J(t)|^2} & (1-p)(1-|s_J(t)|^2)
\end{pmatrix},
\end{equation}
with $p$ and $s_J(t)$ given after \eqref{22} and in \eqref{thesj.1}, respectively. Note that the reservoir entanglement spectrum is entirely determined by \eqref{17.1}.

\begin{thm}[Energy conserving $X$-interaction]
	\label{lem04}
	Consider the system with the $X$-interaction, \eqref{001} for a fixed, finite $N$. 
	\begin{itemize}
		\item[ {\rm \bf (A)}] Let $\rho_\s(t)$ be the reduced dimer density matrix. Its diagonal  is time-independent and the off-diagonal is given by 
		\begin{equation}
		[\rho_{\rm S}(t)]_{12} = \e^{-\i\Omega t}s_{[0,\infty)}(t)    \, \exp\big\{ 4\i\lambda\, {\rm Im}\,\textstyle \sum_{j=1}^N   \bar\alpha_j g_j \tfrac{1-\e^{\i\omega_j t}}{\omega_j} \big\}\ [\rho_{\rm S}(0)]_{12}\, ,
		\label{23}
		\end{equation}
where $s_{[0,\infty)}(t)$ is given in \eqref{thesj.1} ($X$-interaction) with $J=[0,\infty)$.
		\item[ {\rm \bf (B)}]  Let $\rho_J(t)$ be the reduced density matrix of all oscillators having frequencies inside $J$. The rank of $\rho_J(t)$ is two ($t>0$) and in a suitable orthonormal basis, it has the matrix representation \eqref{17.1}.
	\end{itemize}
\end{thm}
The basis in which $\rho_J(t)$ takes the form \eqref{17.1} depends on time $t$, the number $N$ of oscillators and the frequency window $J$. 
  For the energy conserving $D$-interaction ($V=\mu=0$) we denote the reduced dimer and oscillator density matrices by $\rho^0_\s(t)$ and $\rho_J^0(t)$. The superscript $0$ indicates that these are the unperturbed states, as later on, we will switch on the energy exchange interactions. 
\begin{thm}[Energy conserving $D$-interaction]
	\label{lem4'}
Let $N$ be fixed and finite. 
	Denote by $\rho^0_\s(t)$ and $\rho_J^0(t)$ the reduced system and oscillator density matrices evolving according to the $D$-interaction, \eqref{0032}, for $V=\mu=0$.  
	\begin{itemize}
		\item[ {\rm \bf (A)}] The diagonal of $\rho^0_\s(t)$  is time-independent and the off-diagonal  is 
		\begin{equation}
		[\rho^0_{\rm S}(t)]_{12} = \e^{-\i\Omega t}\bar s_{[0,\infty)}(t)  \,[\rho_{\rm S}(0)]_{12}\, ,
		\label{23'}
		\end{equation}
where $\bar s_{[0,\infty)}(t)$ is the complex conjugate of $s_{[0,\infty)}(t)$,  given in \eqref{thesj.1} ($D$-interaction) with $J=[0,\infty)$.
		\item[ {\rm \bf (B)}] The rank of $\rho^0_J(t)$ is two ($t>0$) and in a suitable orthonormal basis, it has the matrix representation \eqref{17.1}. 
	\end{itemize}
\end{thm}

\subsection{$D$-interaction: partial decoherence}

As mentioned in Section \ref{contmodesect}, irreversible dynamics emerges in the continuous mode limit only. An example is the evolution of coherences, encoded in the time dependence of $s_J(t)$, \eqref{thesj.1}. For the $D$-interaction model, we obtain in the continuous mode limit \eqref{25},
$$
|s_J(t)| = \e^{-\int_J h_D(\omega)d\omega} \ \e^{{\rm Re} \int_J h_D(\omega)\, \e^{2\i t\lambda g(\omega)}d\omega}\longrightarrow
\e^{-\int_J h_D(\omega)d\omega},\qquad t\rightarrow\infty,
$$ 
where 
\begin{equation}
\label{hd}
\qquad h_D(\omega) = \tfrac{dn(\omega)}{d\omega} \  |\alpha(\omega)|^2.
\end{equation}
The convergence is due to fast oscillations, causing $\int_J h_D(\omega)\e^{2\i t\lambda g(\omega)}d\omega\rightarrow 0$ as $t\rightarrow \infty$, a process we can control by a `Riemann-Lebesgue Lemma' argument as follows.
For definiteness,  consider $h$ to be a function of $\omega\in[0,\omega_c]$, where $\omega_c$ is some cutoff (at and beyond which $h$ and $h'$ vanish) and suppose that the coupling function $g$ is invertible on $\omega\in[0,\omega_c]$. Using that $\e^{-2\i t\lambda \omega} = \frac{1}{-2\i t\lambda} \partial_\omega \e^{-2\i t\lambda\omega}$ and integrating by parts, we expand
\begin{eqnarray}
\int_0^\infty h(\omega) \e^{-2\i t\lambda g(\omega)} d\omega& =& \int_{g(0)}^{g(\omega_c)} \frac{h(g^{-1}(\omega))}{g'(g^{-1}(\omega))}\   \e^{-2\i t\lambda \omega} d\omega \nonumber\\
&=&\frac{\i}{2t\lambda} \frac{h(0)}{g'(0)} +\frac{1}{4t^2\lambda^2}  \big(\frac{h(g^{-1}(\omega))}{g'(g^{-1}(\omega))}\big)'\,  \big|_{\omega=g(0)}  +O\big((t\lambda)^{-3}\big).   \qquad 
\label{61}
\end{eqnarray}
It follows from \eqref{61} and \eqref{23'} that $|[\rho^0_\s(t)]_{12}|$ decays partially (but not monotonically in time) over the {\em partial decoherence time}
\begin{equation}
\label{57}
t_{\rm pd} =\frac{\sqrt{|\xi|}}{\lambda},
\end{equation}
with $\xi$ given by 
\begin{equation}
\label{58}
\xi =   \Big(\frac{h_D(g^{-1}(\omega))}{g'(g^{-1}(\omega))}\Big)'\,  \Big|_{\omega=g(0)}.
\end{equation}
We point out that the decoherence is only partial, since 
\begin{equation}
\label{56}
\big|[\rho^0_\s(t)]_{12}\big| = \big|[\rho_\s(0)]_{12}\big| \ \e^{-\int_0^\infty h_D(\omega) d\omega}\ \big( 1+ O(t_{\rm pd}^2/t^2)\big)
\end{equation}
does not vanish as $t\rightarrow\infty$. In contrast, we will show that for the $X$-interaction model, we have {\em full decoherence} (see Proposition \ref{lem5}). 

\subsection{$D$-interaction: almost energy conserving regime}

\label{almostsect}

Even for $V$ and $\mu$ small (not vanishing), the energy exchange interaction terms in $H_D$, \eqref{0032}, have a big influence over long time scales. For instance, they are responsible for the process of relaxation (thermalization, if the reservoir is in a thermal state). We consider here time scales of the order $t_{\rm pd}$, \eqref{57}, over which partial decoherence happens and we show that the effects of the non-energy conserving terms are merely a perturbation of this dynamics for small $V, \mu$. We call this the {\em almost energy conserving regime}. 

In the result below, we use the notation $o(x)$ for a real number  s.t. $\lim_{x\rightarrow 0} o(x)/x =0$. Also, we denote by $\|\cdot\|_1$ the {\em trace norm}, namely, $\|A\|_1={\rm Tr} |A|$, where $|A|=\sqrt{A^*A}$ is the absolute value of an operator $A$.
\begin{thm}[Energy exchange $D$-interaction]
	\label{thm01}  
	Let $N$ be finite and consider the $D$-interaction with $\mu=0$ and $V\ge 0$. Suppose that $V/\Omega <\!\!<1$, $V/\lambda <\!\!<1$ and that $\Omega/\lambda <\!\!<  \min_{1\le j\le N} g_j$. Then, for times $0\le t\le t_{\rm pd}$, we have the following.
	\begin{itemize}
\item[{\bf (A)}] The dimer density matrix satisfies
		\begin{eqnarray}
		\label{052}
		\big\| \rho_\s(t) - \rho_\s^0(t) - \tfrac{V}{\Omega}\, \e^{-\sum_{j=1}^N |\alpha_j|^2}\rho^1_\s(t)\big\|_1 = o(V/\Omega),
		\end{eqnarray} 
		where $\rho_\s^0(t)$ is given in \eqref{23'} and 
		\begin{equation}
		\label{070.1}
		\rho_\s^1(t) = 
		\begin{pmatrix}
		{\rm Re} \ \big(1-\e^{-\i \Omega t}\big) \ [\rho_\s(0)]_{12}  & (p-\tfrac12)(1-\e^{-\i \Omega t})\\
		(p-\tfrac12)(1-\e^{\i \Omega t})& -{\rm Re} \ \big(1-\e^{-\i \Omega t}\big) \ [\rho_\s(0)]_{12}
		\end{pmatrix}.
		\end{equation}
		The remainder is uniform in $t$ for $0\le t\le t_{\rm pd}$. 
		
\item[{\bf (B)}] The oscillator density matrix satisfies
		\begin{equation}
		\big\| \rho_J(t) -\rho^{00}_J(t)  - \tfrac{V}{\Omega}\,  \e^{-\sum_{\{ j\, :\, \omega_j\in J\}} |\alpha_j|^2} \rho_J^1(t)\big\|_1 =o(V/\Omega),
		\end{equation}
		where $\rho_J^{00}(t)$ is the $3\times 3$ matrix obtained by adding a (third) row and a (third) column consisting of zeroes to the matrix in \eqref{17.1}, and, expressed in the same basis, the hermitian  $\rho_J^1(t)$ has matrix elements
		\begin{eqnarray}
		{} [\rho_J^1(t)]_{11} =  - [\rho_J^1(t)]_{22} &=& 2\, {\rm Re}\, \bar v_J(1-s_J)\nonumber\\
		{} [\rho_J^1(t)]_{33} &=& 0\nonumber\\
		{} [\rho_J^1(t)]_{12} &=& (v_J-\bar v_J s_J) \tfrac{1-s_J}{\sqrt{1-|s_J|^2}}-v_J\sqrt{1-|s_J|^2} \nonumber\\
		{} [\rho_J^1(t)]_{13} &=& (\delta_J)^{-1/2}\, {\cal N}_J (v_J-s_J\overline v_J)\nonumber\\
		{}[\rho_J^1(t)]_{23} &=& -\sqrt{1-|s_J|^2} \,\overline v_J\, (\delta_J)^{-1/2} \, {\cal N}_J.
		\label{rho1}
		\end{eqnarray}
		The parameters are given by 
		\begin{eqnarray}
		\delta_J &=& \e^{-\sum_{\{j\, :\, \omega_j\in J \}}|\alpha_j|^2 }\label{-01}\\
		v_J (t)\equiv v_J&=& \tfrac12 (1-\e^{-\i \Omega t})\  [\rho_\s(0)]_{12}\  \e^{-\sum_{\{ j\, :\, \omega_j \not\in J\}} |\alpha_j|^2}\label{-02}\\
		{\cal N}_J  &=& \sqrt{1-2\delta_J\, \tfrac{1-{\rm Re}\, s_J}{1-|s_J|^2}}\ .
		\label{-03}
		\end{eqnarray}
	\end{itemize}
\end{thm}
We refer to Theorem \ref{thm1a} below for results when in addition, $\mu>0$.

\section{Dynamics in the continuous mode limit}
\label{sect3}

\subsection{Dynamics of the dimer and of $s_J$}

In the continuous mode limit \eqref{25}, the discrete mode expressions for $s_J(t)$, \eqref{thesj.1}, become \eqref{thesj}.

\begin{prop}[$X$-interaction]
	\label{lem5} 
	The dynamics of $\rho_\s$ and $s_J$ for the $X$-interaction model, in the continuous mode limit, satisfies the following. In what follows, $o_t$ denotes a real number satisfying $\lim_{t\rightarrow\infty} o_t=0$.
	\begin{itemize}
		\item[{\bf (A)}] Suppose that $h_X(\omega)$, \eqref{theh}, is continuous as $\omega\rightarrow 0_+$.
		Then 
		\begin{equation}
		\label{30}
		\big|\, [\rho_{\rm S}(t)]_{12}\, \big|  =  \big|\, [\rho_{\rm S}(0)]_{12}\, \big|\ \e^{-2\pi t  \lambda^2 h_X(0)[1+o_t]}.
		\end{equation}		
		\item[\bf (B)] Let $J=[0,\omega_1]$, where $0<\omega_1\le\infty$. Then for all $t\ge 0$,
		\begin{equation}
		\label{29}
		|s_J(t)| = \e^{-2\pi t \lambda^2 h_X(0)\, [1+o_t]}.
		\end{equation}
		\item[\bf (C)] Let $J=[\omega_0,\omega_1]$, where $0<\omega_0<\omega_1< \infty$ and set $m_J=\inf_{\omega\in J} h_X(\omega)$, $M_J = \sup_{\omega\in J} h_X(\omega)$. We have for all $t\ge 0$,
		\begin{equation}
		\label{28}
		\e^{-4\lambda^2 M_J\, \frac{\omega_1-\omega_0 +2/t}{\omega_0^2}} \le  |s_J(t)| \le \e^{-4\lambda^2 m_J\, \frac{\omega_1-\omega_0-2/t}{\omega_1^2}}.
		\end{equation}
	\end{itemize}	
	
\end{prop}
According to (A),  the dimer undergoes full decoherence for couplings $h_X(0)\neq 0$, exponentially quickly in time.
\begin{prop}[$D$-interaction, $V=\mu=0$] 
	\label{prop2} 
	The dynamics of $\rho_\s$ and $s_J$ for the energy conserving $D$-interaction model, in the continuous mode limit, satisfies the following. 
	\begin{itemize}
		\item[{\bf (A)}] We have 
		\begin{equation}
		\label{056}
		\big| [\rho^0_\s(t)]_{12}\big|   =   \big| [\rho_\s(0)]_{12}\big|  \   \e^{-\int_0^\infty h_D(\omega) d\omega}\ \Big( 1+ O\big( t^2_{\rm pd}/t^2\big)\Big),
		\end{equation}	
		where $t_{\rm pd}\propto 1/\lambda$ is the partial decoherence time (c.f. \eqref{57}).
		
		\item[{\bf (B)}] For any $J\subseteq [0,\infty)$, the quantity $s_J(t)$, given in \eqref{thesj}, satisfies
		\begin{equation}
		\label{itself}
		s_J(t)=  \e^{-\int_J h_D(\omega) d\omega} 
		\ \Big( 1+ O\big( t^2_{\rm pd}/t^2\big)\Big).
		\end{equation}
	\end{itemize}
\end{prop}
The expansions are correct provided $t_{\rm pd}>0$, which translates into a condition of effective coupling, see \eqref{62}, \eqref{60}. The dimer undergoes partial decoherence.

\bigskip

We close the discussion by explaining how the energy-exchange term $\propto V$ of the $D$-interaction model, \eqref{0032}, influences the entanglement entropy. For some results including additionally the term $\propto \mu$, see Section \ref{dintsect}. 

\begin{prop}[Entanglement entropies, $D$-interaction, $V>0$]
\label{lem4}
Consider the dynamics associated with the $D$-interaction Hamiltonian, for $V>0$ (and $\mu=0$), in the continuous mode limit. Define
	\begin{equation}
	r_J^1(t) = 2(1-2p)\tfrac{\, \e^{-\int_{{\mathbb R}} h_D(\omega) d \omega}}{r_J(t)} \ {\rm Re}\, [\rho_\s(0)]_{12} (1-\e^{-\i t\Omega})(1-\bar s_J(t)).
	\label{rjone}
	\end{equation}
Suppose that $V/\Omega <\!\!<1$, $V/\lambda <\!\!<1$ and that $\Omega/\lambda <\!\!< \inf_{\omega\ge 0} g(\omega)$. Then, for times $0\le t\le t_{\rm pd}$, we have:
\begin{itemize}
\item[{\rm \bf (A)}] The eigenvalues of $\rho_\s(t)$ are
	\begin{equation}
		{\rm spec}(\rho_\s(t)) = \big\{ \tfrac12 \pm \tfrac12\big[ r_{[0,\infty)}(t) -\tfrac{V}{\Omega} r^1_{[0,\infty)}(t)\big] \big\} +o(V/\Omega).
		\label{87} 
	\end{equation}

\item[{\rm \bf (B)}] The eigenvalues of $\rho_J(t)$ are, in the continuous mode limit,
	\begin{equation}
	{\rm spec}(\rho_J (t)) = \big\{ \tfrac12 \pm \tfrac12\big[ r_J(t) -\tfrac{V}{\Omega} r^1_J(t)\big]\big\}\cup\{0\}  +o(V/\Omega).
	\label{87'} 
	\end{equation}
\end{itemize}
The small-o notation means that $o(x)$ is a real number  s.t. $\lim_{x\rightarrow 0} o(x)/x=0$.  Also, $r_J(t)$ is given in \eqref{rj}. 
\end{prop}

{\bf Discussion. } {\bf 1.} Theorem \ref{thm01} and Proposition \ref{lem4} show that the energy-exchange perturbation $\propto V$ in \eqref{0032} leaves the rank of the oscillator density matrices $\rho_J(t)$ unchanged, up to order $V/\Omega$ (see \eqref{87'}).

{\bf 2.} For $p=1/2$, the term on the right side of \eqref{87}, which is linear in $V/\Omega$, vanishes. The entanglement entropy of the dimer decreases if the gap between the two eigenvalues of $\rho_\s(t)$ increases, which happens exactly if the term $\propto V/\Omega$ in the parentheses in \eqref{87} has a positive sign. This shows that the energy-exchange interaction produces a correction to the (strictly positive) entanglement entropy caused by the energy conserving interaction, whose sign fluctuates in time and depends on the initial dimer state.

\section{Energy exchange, dissipation, equilibration}
\label{ultimsect}

\subsection{Systems close to thermal equilibrium} 
\label{secequil}

The theory of open quantum systems analyzes the dynamics of `small systems' coupled to `environments' (also called reservoirs or quantum noises). The environment itself is a quantum system but it shows irreversible dynamical effects, both in its own evolution and in that of a small system coupled to it. A reservoir is obtained by a limiting procedure from a `normal' (finite) quantum system, for instance by taking a continuous mode limit. The limits turn purely discrete energy spectrum of finite systems into continuous spectrum, which causes irreversibility in the dynamics. It is plausible to expect that a system with continuous energy (frequency) spectrum shows {\em dissipative dynamics},  suggesting that reservoirs have {\em a unique stationary state}. The same thing happens when a transition of discrete to continuous frequencies is caused by a thermodynamic limit, where the volume of a reservoir increases, see \cite{JM,MMLecturenotes,BR}. 

A non-interacting system environment complex, described by a Hamiltonian $H_0= H_\s +H_\r$, has {\em a manifold of stationary states}, spanned by $|\psi_E\rangle\langle\psi_E|\otimes\rho_\r$, where $\psi_E$ are the system eigenfunctions and $\rho_\r$ is the reservoir invariant state. What happens as the system and reservoir start to interact, due to a term $H_{\s\r}$ in the Hamiltonian? In the situation where the reservoir is in (or close to) thermal equilibrium (the thermodynamic limit of finite-volume Gibbs states), one can show that, for very generic interactions $H_{\s\r}$, the interacting system has a {\em unique stationary state}, which is the coupled (interacting) equilibrium state. The fact that such a coupled equilibrium exists is guaranteed by the {\em stability theory of KMS states}.\footnote{For continuous modes (frequencies) systems, the Hamiltonian $H_\r$ has continuous spectrum and one does not know how to interpret the candidate $\e^{-\beta H_\r}$ for an equilibrium density matrix, since its trace is infinite. A better characterization of equilibrium is the KMS characterization (Kubo-Martin-Schwinger) \cite{BR}, which reduces to the usual Gibbs state formulation for equilibrium in finite systems.} This theory says, quite generally, that if an unperturbed Hamiltonian $H_0$ generates a dynamics which has an equilibrium (KMS) state $\rho_0$  then a perturbation $H_0+H_{\s\r}$ generates a dynamics with associated equilibrium state $\rho_{\s\r}$, which varies continuously in $H_{\s\r}$. The fact that $\rho_{\s\r}$ is the {\em only} stationary state can be proven for many open systems, provided an efficient coupling condition is satisfied. The latter requires that {\em energy exchange} takes place between the system and the environment, and it will not be satisfied in general for models where the system energy is conserved.\footnote{Suppose $H=H_\s+H_\r+H_{\s\r}$ is energy conserving, meaning that $[H_\s,H_{\s\r}]=0$. Then all initial states $|\psi_E\rangle\langle\psi_E|\otimes\rho_\r$, where $H_\s\psi_E = E\psi_E$ and $\rho_\r$ is an arbitrary reservoir density matrix produce stationary system states, ${\rm Tr}_\r \{ \e^{-\i tH} ( |\psi_E\rangle\langle\psi_E|\otimes\rho_\r) \e^{\i tH}\} = |\psi_E\rangle\langle\psi_E|$. This observation hints at the need of energy exchange to guarantee uniqueness of a stationary state for the whole, coupled system-reservoir complex.} We sum up the {\em situation close to equilibrium} as follows:
\begin{itemize}
\item[1.] The coupled system-reservoir complex has an equilibrium state.

\item[2.] Due to efficient coupling, requiring in particular energy-exchange, the coupled equilibrium is the only stationary state.

\item[3.] The coupled dynamics is dissipative and drives any initial state towards the coupled equilibrium. 
\end{itemize}
For small $H_{\s\r}$ the reduction of the coupled equilibrium to the system alone is the usual Gibbs state $\propto  \e^{-\beta H_\s}$ (modulo $O(H_{\s\r})$ terms) at the temperature inherited from the reservoir. For strong system-reservoir coupling, the reduced state is a Gibbs state (again at inverse temperature $\beta$) with respect to a Hamiltonian having {\em renormalized energies} \cite{Merkli1}. The renormalizations are $\propto (H_{\s\r})^2$ and can be large. In either case, the small system becomes {\em equilibrated} due to the contact with a thermal reservoir, in the course of time.

We have developed the dynamical resonance theory \cite{Merkli2}, allowing a rigorous proof of 1.-3., for initial states of the form $\rho_\s\otimes\rho_\r$ (or perturbations thereof), where $\rho_\r$ is the reservoir equilibrium, hence a {\em mixed} state. However, for the analysis of entanglement entropy, one has in mind {\em pure} system-reservoir initial states. A modification at the very core of the resonance theory is necessary to make it applicable to the study of entanglement entropy. While we are working to expand the theory in this direction, the present paper investigates the almost energy conserving situation only. 

\subsection{Emergence of Gibbs states from reducing pure states} 
\label{subsubgibbs}
In the ``standard setup of open systems''  described in Section \ref{secequil}, the starting point is a state where the reservoir is in (or close to) equilibrium. A different question is how a system equilibrium (Gibbs) state emerges due to the interaction with a reservoir when initially the system-reservoir state is {\em pure}. 

Suppose $H=H_\s+H_\r+H_{\s\r}$ where the reservoir has discrete, but close lying energy levels, spaced by maximally $\Delta B$, while the system energy spacing is $\Delta \epsilon$. Denote by $\rho(B)$ the reservoir density of states and define the effective temperature by $\beta\equiv\beta(B) = \tfrac{d}{d B}\ln \rho(B)$. In the regime $\Delta B <\!\!< \| H_{\s\r}\|<\!\!<\Delta\epsilon$ (and under additional assumptions), it is shown in \cite{Tasaki, Tasakinote} that for an initial pure state $\Phi(0)$ of the system plus reservoir, having a (coupled) energy distribution peaked at a value $E$, one has 
\begin{equation}
\label{TasakiResult}
\langle A\rangle_t \equiv \langle \Phi(t), (A\otimes\bbbone_\r)\Phi(t)\rangle \approx \langle A\rangle_{\beta(E)}^{\rm can},
\end{equation} 
for all system observables $A$, where $\Phi(t)=\e^{-\i tH}\Phi(0)$ and $\langle A\rangle_{\beta}^{\rm can} = {\rm Tr}_\s(\e^{-\beta H_\s} A)/{\rm Tr}_\s\e^{-\beta H_\s}$ is the system Gibbs canonical equilibrium state. Relation \eqref{TasakiResult} is valid for sufficiently large, typical times $t$ (it does not hold for all times, since the reservoir is not taken to be in the continuous mode limit and hence the overall dynamics is quasi-periodic). It would be interesting to give a proof of \eqref{TasakiResult} for our $X$-interaction model with a coherent state initial reservoir, especially for strong system-reservoir couplings (relevant, for instance, in quantum biological processes). In view of the techniques developed in \cite{Merkli1}, we might expect such a result to hold, with $H_\s$ replaced with a renormalized Hamiltonian.

The following two arguments are often used to heuristically derive the emergence of the system Gibbs state, obtained by reducing a full (pure state) density matrix.
\begin{itemize}
	\item[1.] (Decoherence in the energy basis.) After the system and reservoir have interacted for some time, the system will be in a state which is {\em diagonal in its (uncoupled) energy basis}, and which {\em does not depend on the initial system state}. It follows from this (assumption, or principle) that the system state is entirely described by the populations (occupation probabilities) $p_i$ of the energy levels, and that those populations are functions of the energy only, $p_i= f(E_i)$. It is implicitly assumed here that the system and reservoir exchange energy since if they were not, then the system populations would be time-independent and they would not settle to values independent of the initial system state.
	\item[2.] When coupling {\em two identical, independent} systems to the same reservoir (no direct interaction), the population probabilities of the joint two-system complex should split into products of single system probabilities, due to their independence. Here, it is assumed that the systems-reservoir interaction is weak, so that independence of the two systems is approximately preserved even while they are in contact with the same reservoir. This leads to the constraint $f(x+y)=f(x)f(y)$, whose solution has the Gibbs distribution form $f(x)\propto \e^{-\beta x}$, where $\beta$ is a reservoir-dependent constant. 
\end{itemize}
In the arguments of \cite{Tasaki, Tasakinote}, decoherence is derived by observing that for the models considered there, the stationary Schr\"odinger equation of the interacting system can be interpreted as that of a single quantum particle in a potential. A weak system-energy interaction makes this potential vary slowly in space and as a consequence,  eigenstates of the interacting system associated with energies differing by more than the interaction energy have roughly disjoint support (quasiclassical analysis), which leads to decoherence (c.f. equations (6) in \cite{Tasaki}). It is not clear to us if this reasoning is applicable to situations we are interested in here, namely a dimer {\em strongly} coupled to an environment. 

While the above considerations are based on the {\em dynamics}, there is also a `static' link between reduced states of a large system and Gibbs states. In particular, for a collection of solvable many-particle free lattice particle systems and spin chain models in their ground state, the reduced density matrices of parts of the system and their entanglement spectra have been analyzed in \cite{PeschelReview,PeschelEisler,Peschel,ChenFradkin}. The key finding is that the reduced density matrix $\rho_\alpha$ (where $\alpha$ labels a lattice subregion) is {\em of thermal form}, $\rho_\alpha\propto \e^{-H_\alpha}$, where $H_\alpha$ is an effective, {\em local}, {\em free particle} Hamiltonian, but $H_\alpha$ is not the original Hamiltonian restricted to the region $\alpha$. In these models, a thermal structure thus emerges generically by reduction of the ground state to a subregion of space.

\section{Numerical simulations}
\label{numsimsect}

In our numerical simulations we use Eq. (\ref{SJT}) for entanglement entropy of the environment, written as 
\begin{eqnarray}
\label{21a}
S_E(t)&=& -\tfrac12 (1+r(t) )\ln \tfrac12(1+r(t) ) - \tfrac12 (1 - r(t)) \ln \tfrac12(1-r(t))
\label{22a}
\end{eqnarray}
where $r(t) = \sqrt{1 -4p(1-p)(1-|s(t)|^2)}$
and
\begin{equation}
\label{24a}
|s(t)| = \exp\Big(-4\lambda^2\int^\infty_0 \ h(\omega)\frac{1-\cos(\omega t)}{\omega^2} \ d\omega\Big).
\end{equation}
We adopt units in which $\hbar=1$.

\subsection{$\mathbf X$-interaction}
\begin{figure}
\begin{center}
\scalebox{0.34}{\includegraphics{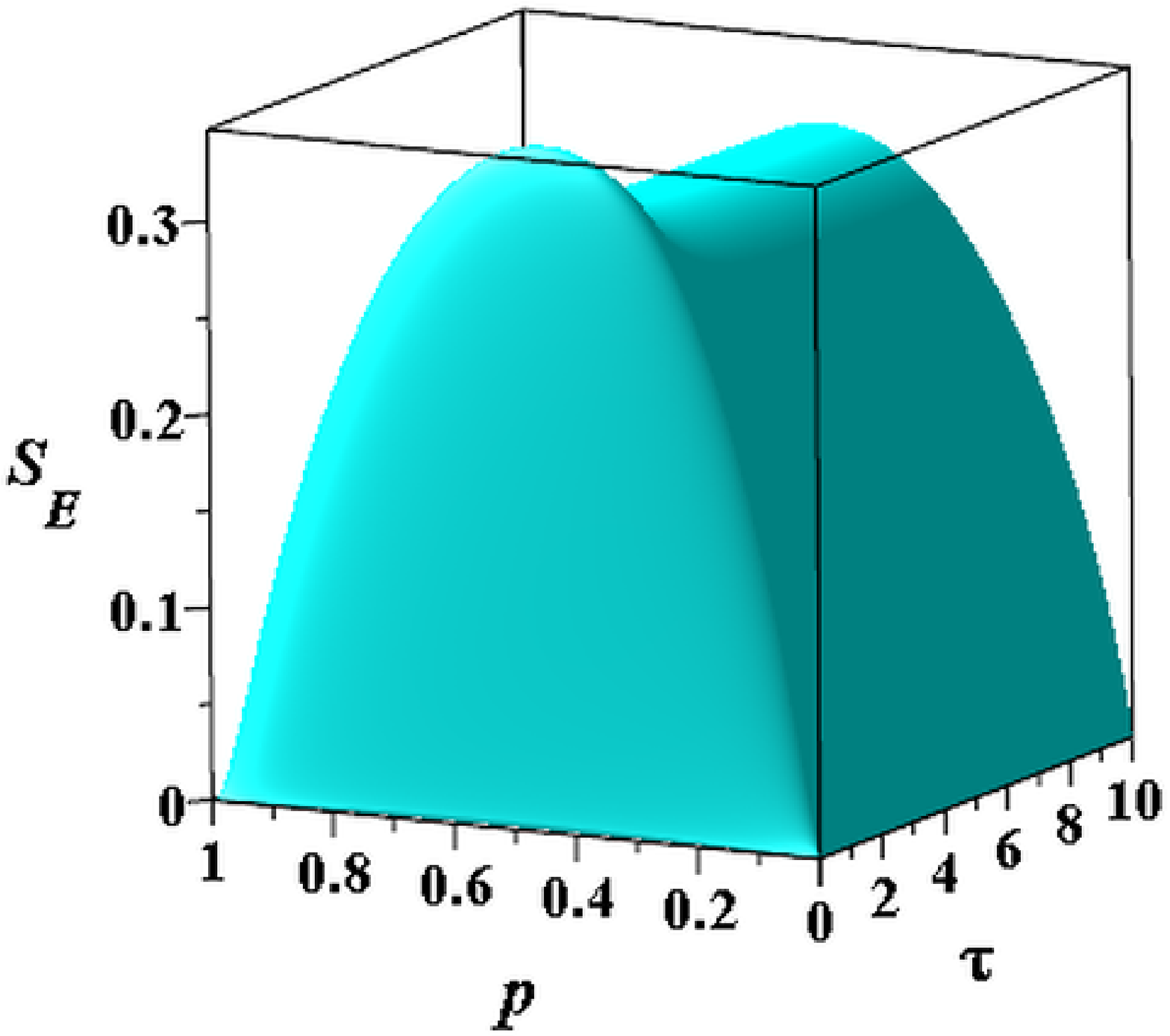}}	
(a)
\scalebox{0.3}{\includegraphics{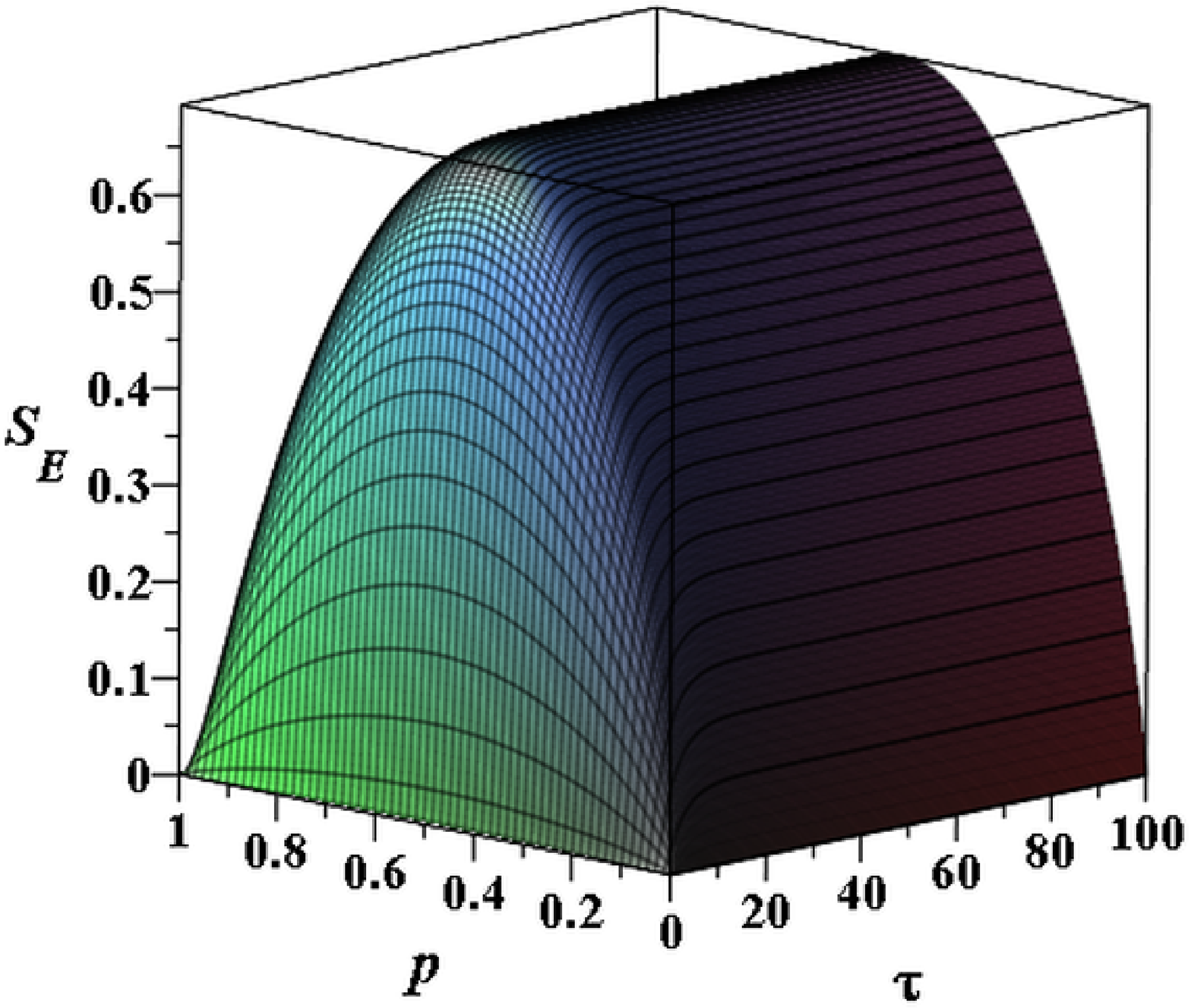}}	
(b)
\end{center}
\caption{(Color online) $X$-interaction. The entanglement entropy  $S_E$ as a 
function of the donor population $p$  and the dimensionless time $\tau = \omega_c t$. (a) $q=1$, $ \varepsilon=0.1$;  (b) $q=-1$,  $\varepsilon=0.1$. Note: the maximal possible value of $S_E$ is $\ln 2\approx 0.69$. 
\label{Fig_1}}
\end{figure}

Let us consider the function $h_X(\omega) \equiv h(\omega)$, given in \eqref{theh}, of the form 
  \begin{equation}	
 \label{e1.1}
 h(\omega) =A_q\,\omega^{2q+2} \e^{-\omega/\omega_c},
 \end{equation}
with an exponential high frequency cutoff $\omega_c>0$, where $A_q>0$ and where $p\in \mathbb R$ determines the low frequency behavior. It is convenient to introduce a new variable, $z = \omega/\omega_c$, and a dimensionless time, $\tau =\omega_c t$. Then, Eq. (\ref{24a}) can be written as,
  \begin{equation}
\label{25a}
|s(t)| = \exp\Big(-\varepsilon\mathcal Q_2(\tau) \Big),
\end{equation}
  where $\varepsilon = 4\lambda^2 A_q \omega_c^{2q+1}$ and 
\begin{align}
\mathcal Q_2(\tau) =\int_0^\infty {z^{ 2 
q} e^{-z}(1-\cos(\tau z))} \d z.
\label{Q2}
\end{align}  
Below, we consider two cases: $q>-1/2$ and $q=-1$. Performing the integration in (\ref{Q2}) gives the following.
 \begin{itemize}
 	\item Case: $q>-1/2$
\begin{align}
\mathcal Q_2(\tau) =\Gamma (2q+1)\Re\bigg( 1-  \frac{1}{(1+ i
		\tau)^{2q+1}}\bigg),
\end{align}  
where $\Gamma(u) = \int_0^\infty \e^{-z}z^{u-1}\d z$ is the gamma function, defined for $u\in\mathbb C$ with $\Re u>0$.

 	\item Case: $q=-1$
 	\begin{align}
\mathcal Q_2(\tau) =\tau \arctan\tau -\tfrac{1}{2}\ln(1+\tau^2).
\label{Q2b}
\end{align}  
 \end{itemize} 
 For both these cases, the EE,  $S_E(\tau)$,  is presented in Fig. \ref{Fig_1}. For $q=1$, we have: $h(0)=0$, and we see from Fig. \ref{Fig_1}a  that only partial decoherence occurs in the X-interaction model. The function $|s(t)|$ does not decay to zero for large times. Correspondingly, the EE $S_E(t)$ does not approach its maximum for a given value of $p$. In particular, for $p=1/2$, $\lim_{\tau\rightarrow\infty}S_E(\tau)\approx 0.3$. At the same time, for $q=-1$ and $p=1/2$, the EE reaches its maximum: $\lim_{\tau\rightarrow\infty}S_E(\tau)=\ln2\approx 0.7$, as in this case  full decoherence takes place ($\lim_{\tau\rightarrow\infty} s(\tau)=0$).

\subsubsection{Contribution of low and high frequencies to the EE}

\begin{figure}[h]
\begin{center}
\scalebox{0.21}{\includegraphics{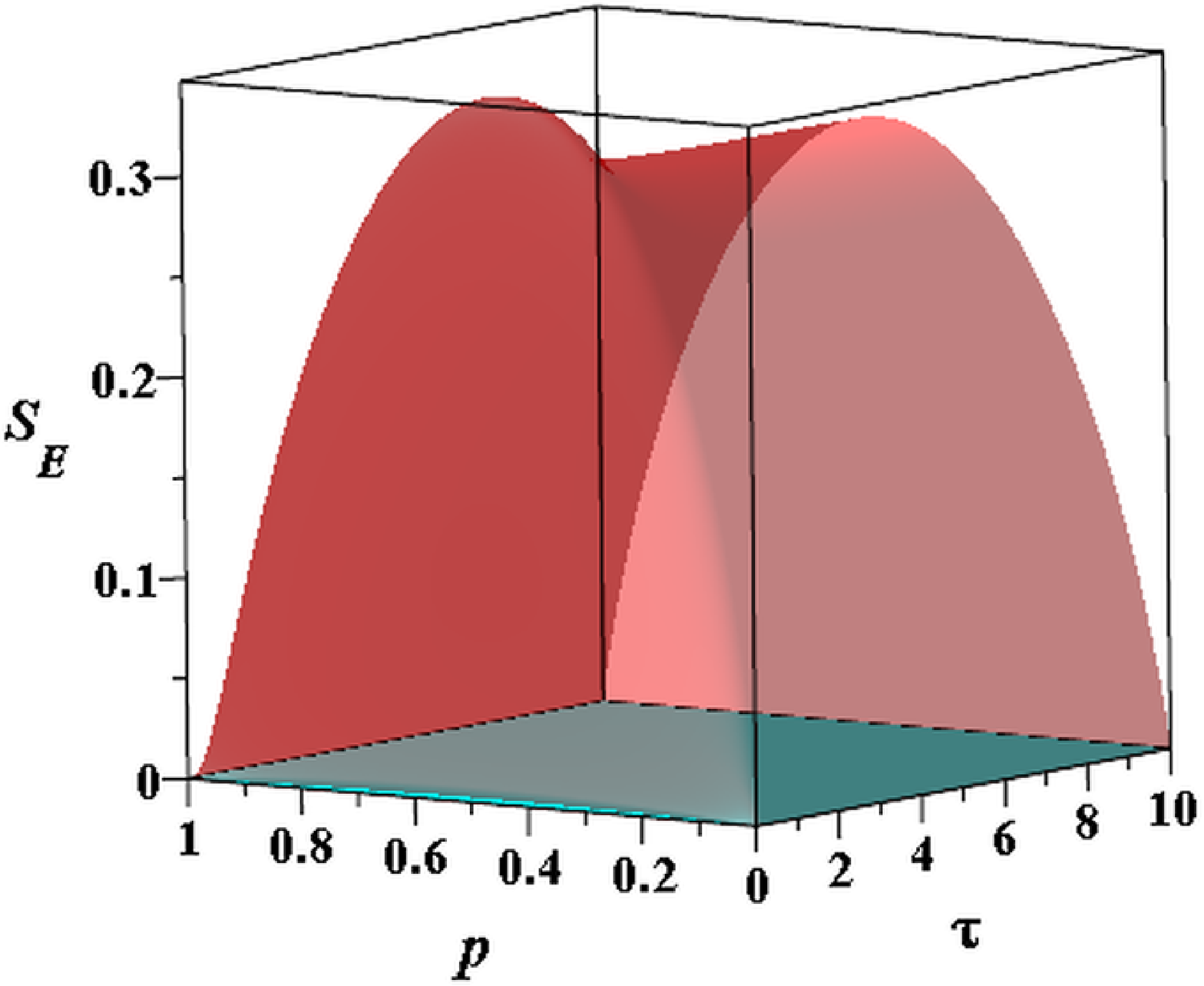}}	
(a)
\scalebox{0.23}{\includegraphics{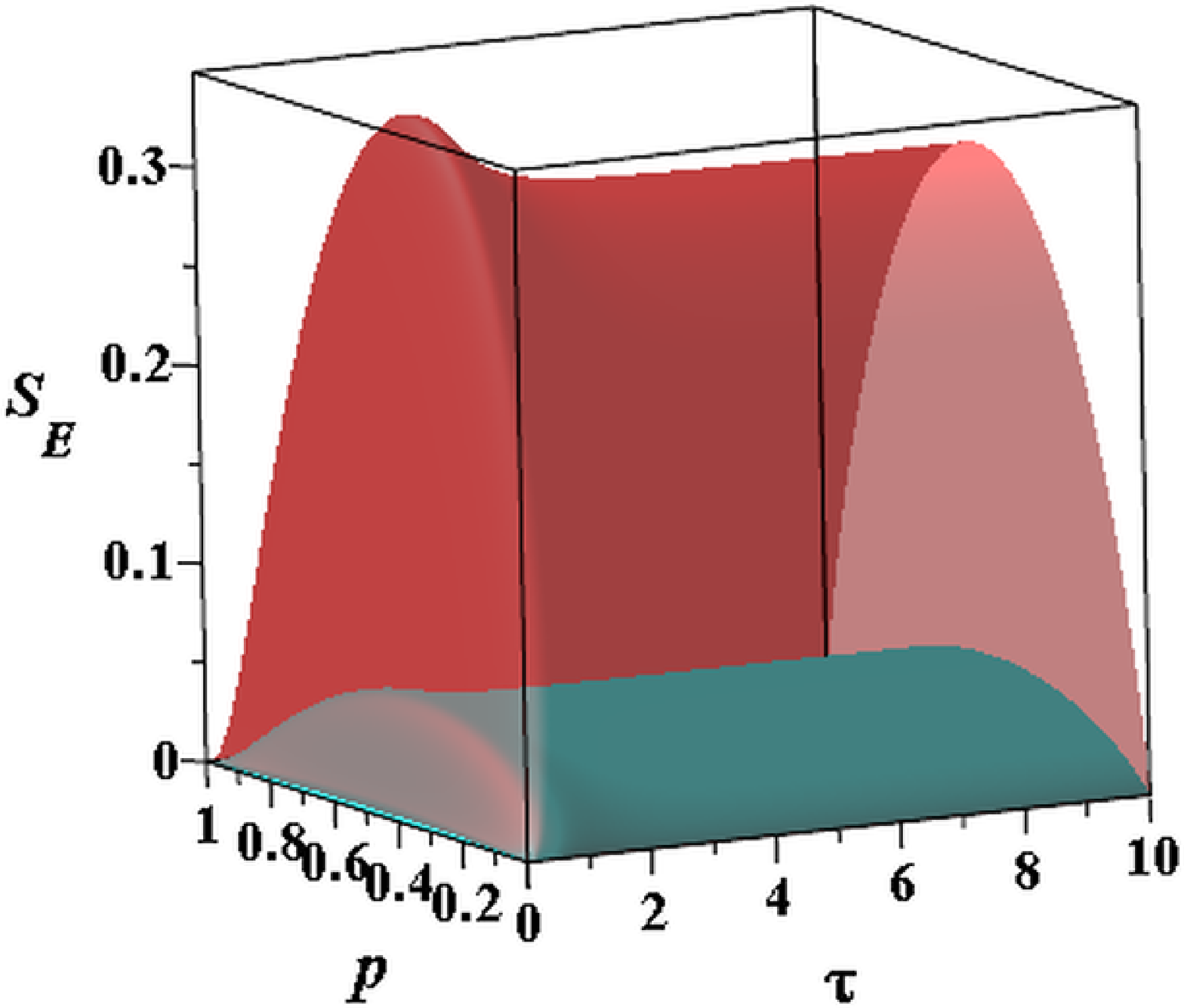}}	
(b)
\scalebox{0.21}{\includegraphics{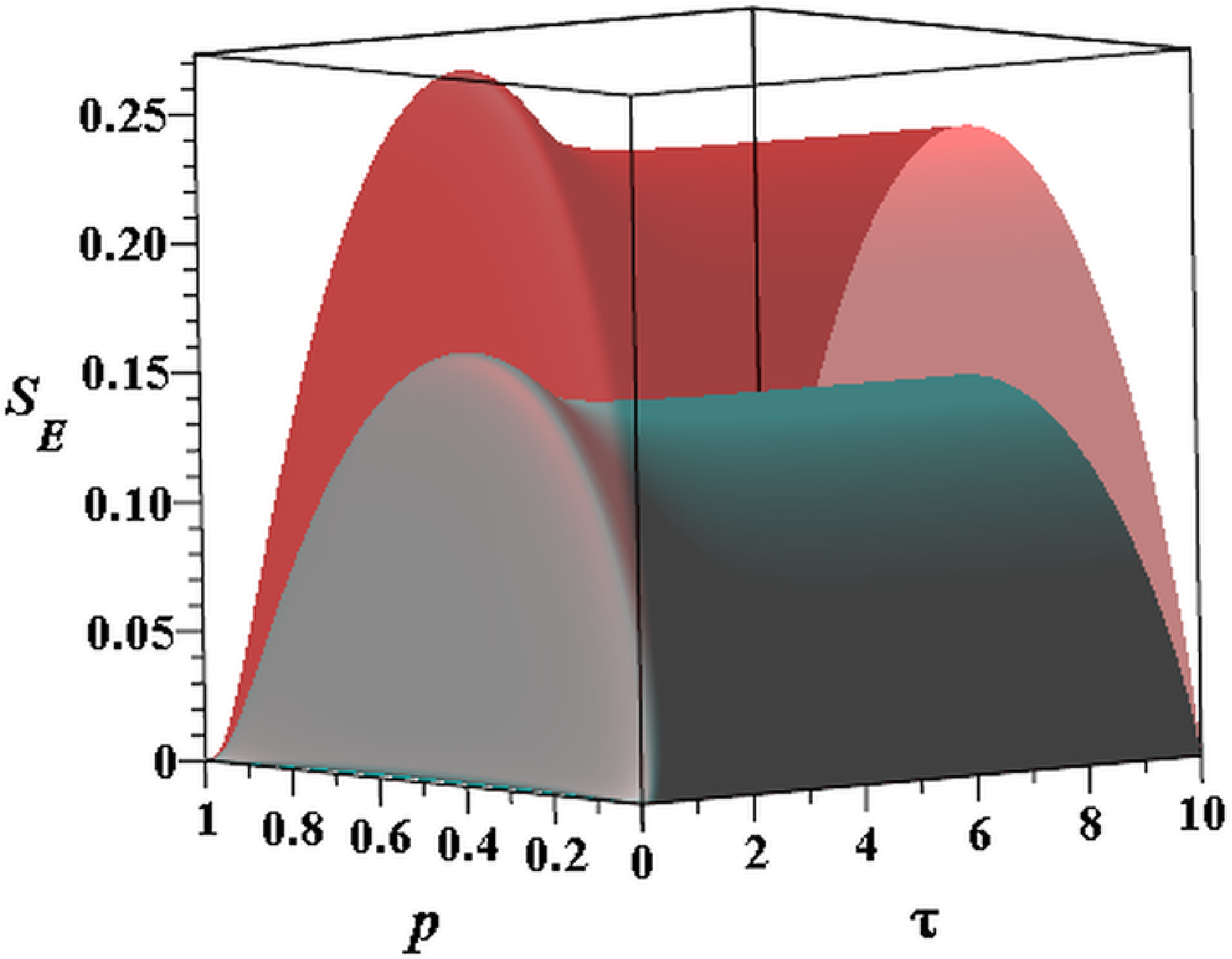}}	
(c)
\scalebox{0.25}{\includegraphics{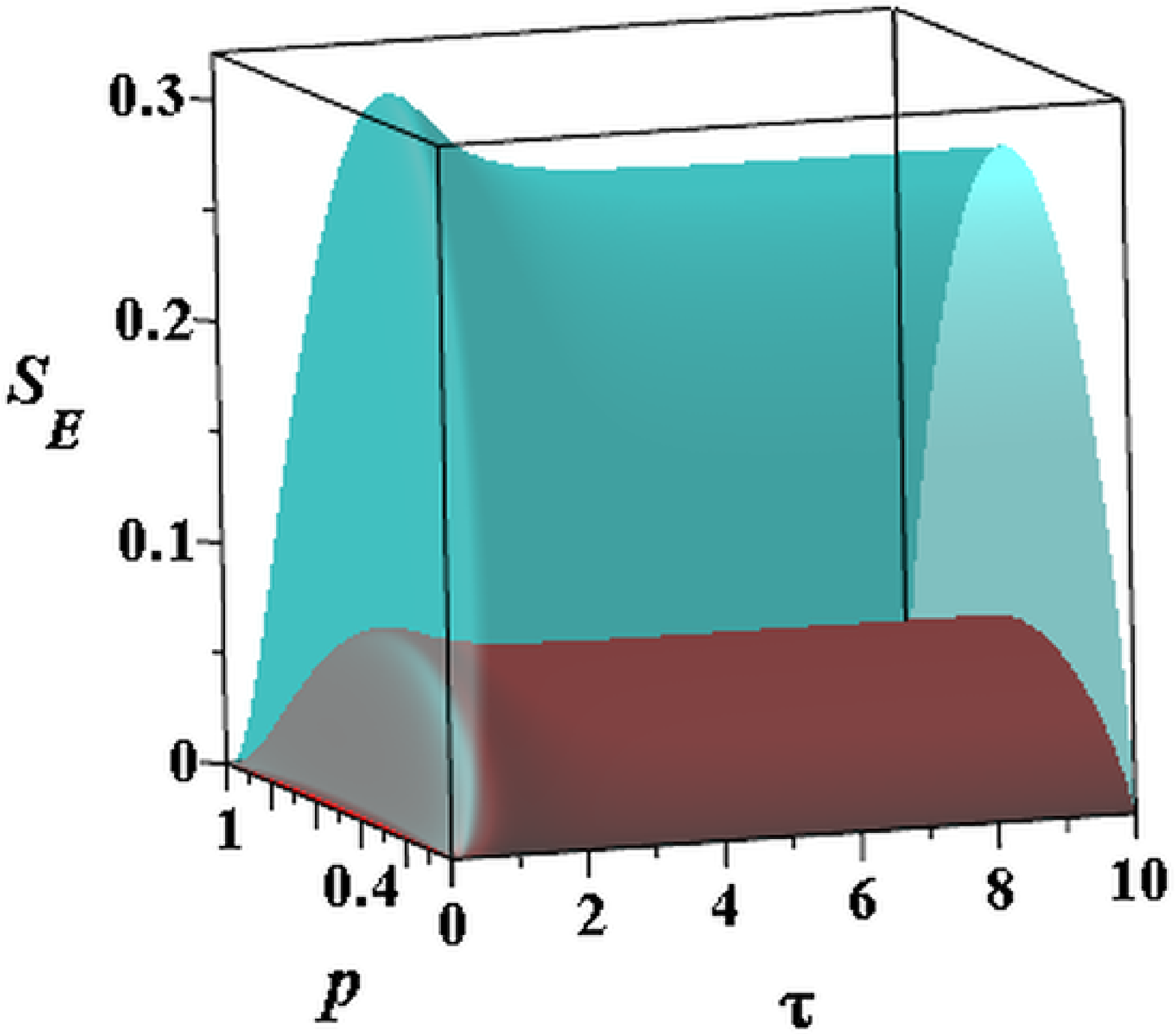}}	
(d)
\end{center}
\caption{(Color online) $X$-interaction. The entanglement entropy,  $S_E$, as a 
function of $p$  and $\tau$:  $q=1$, $ \varepsilon=0.1$. Red surfaces correspond to high frequency contributions $\omega > \omega_0$.  Cyan surfaces give the low frequencies contributions, $\omega < \omega_0$. (a) $z_0 =0.1$, (b) $z_0 =1$, (c) $z_0 =2$. (d) $z_0 =5$. 
\label{Fig_2}}
\end{figure}
To estimate the contribution of low and high frequencies, we consider lower, $\omega_0$, and upper, $\omega_c$, cutoff frequencies in the function $\mathcal Q_2(\tau)$ by introducing,
  \begin{align}
 \mathcal Q_2(\tau,z_0) =\int_0^{z_0} {z^{ 2 
 q} e^{-z}(1-\cos(\tau z))} \d z\quad \mbox{and} \quad \tilde{\mathcal Q_2}(\tau,z_0) =\int_{z_0} ^\infty{z^{ 2 
 q} e^{-z}(1-\cos(\tau z))} \d z,
 \label{Q2ab}
 \end{align}  
where $z_0 = \omega_0/\omega_c$. We have $ {\mathcal Q_2}(\tau) =  {\mathcal Q_2}(\tau,z_0)  +\tilde{\mathcal Q_2}(\tau,z_0)$, but the EE, $S_E(\tau)$, is of course not a linear function of ${\mathcal Q_2}$. In Fig. \ref{Fig_2}  we evaluate $S_E(\tau)$, given in \eqref{22a}, in which the parameter $|s(\tau)|$ is chosen to be either $|s(\tau)| = \e^{-\varepsilon\mathcal Q_2(\tau,z_0)}$ or  $ 
|\tilde s(\tau)| = \e^{-\varepsilon\tilde{\mathcal Q_2}(\tau,z_0)}$ (the low frequency or high frequency contributions to the EE, respectively).

\subsubsection{Contribution of low and high frequencies for $h(\omega)= h_0$ inside of the frequency interval $(\omega_0, \omega_1) $}

\begin{figure}[h]
	\begin{center}
		\scalebox{0.25}{\includegraphics{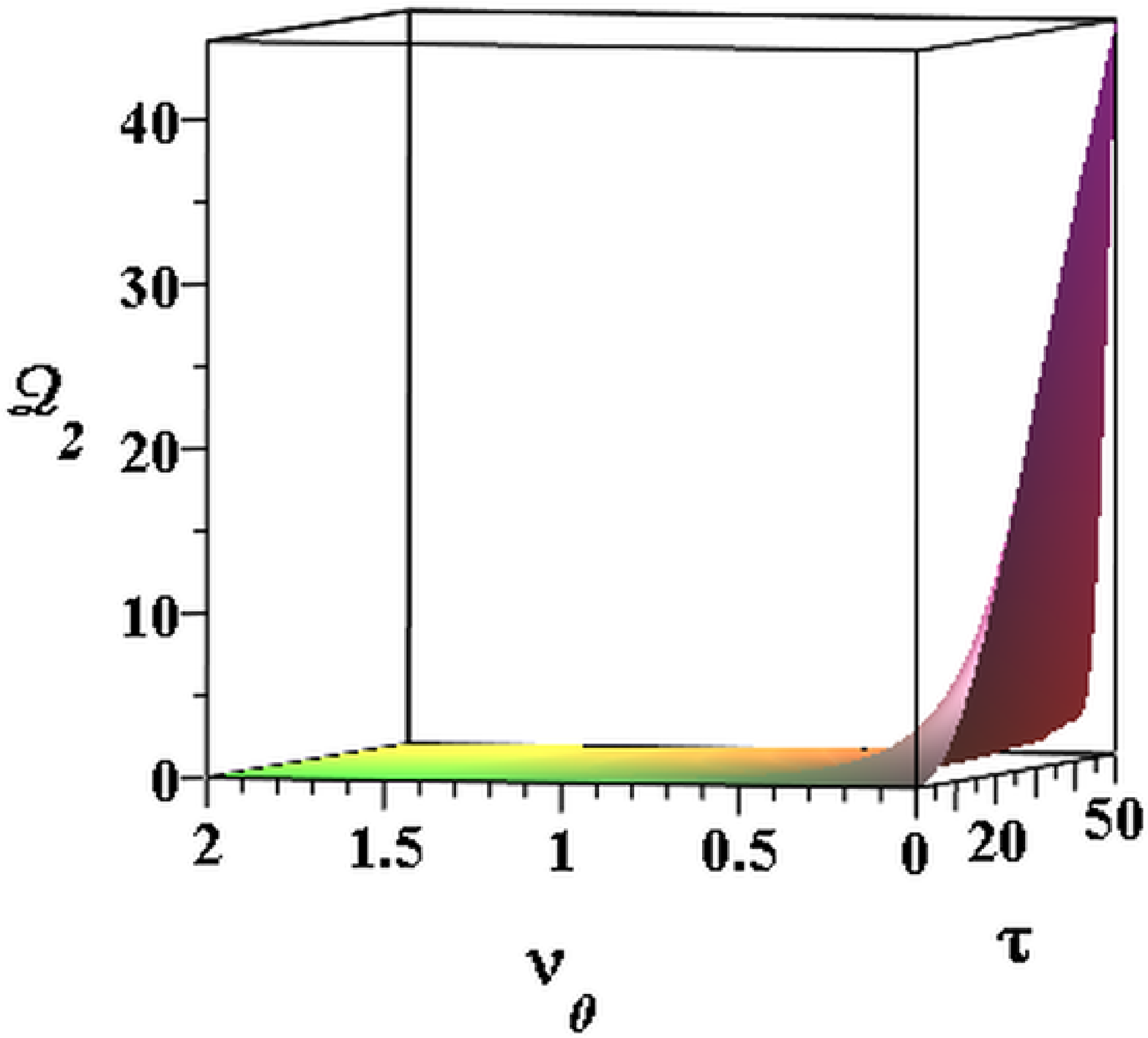}}	
		(a)
		\scalebox{0.25}{\includegraphics{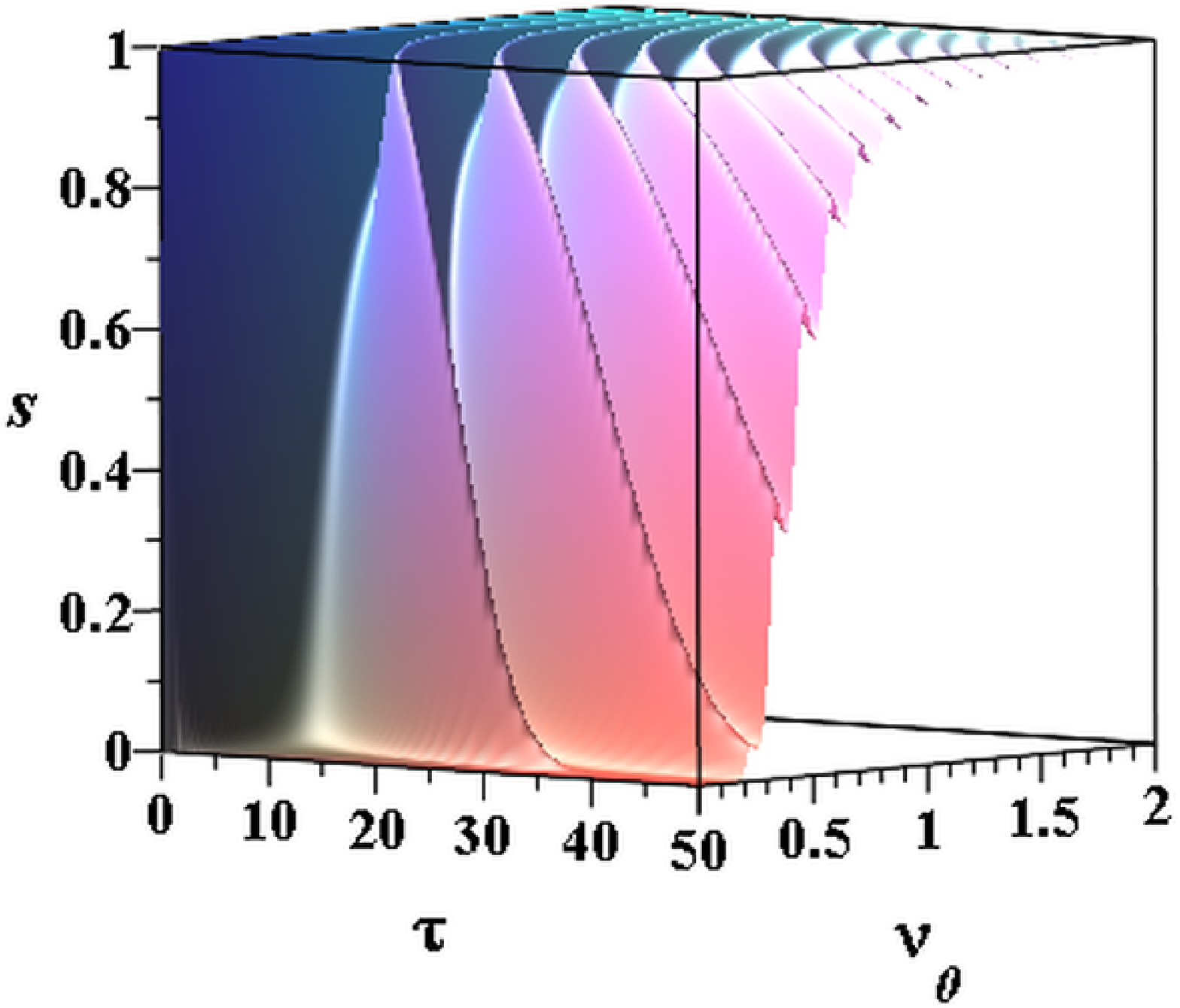}}	
		(b)
	\end{center}
	\caption{(Color online) $X$-interaction. For the fixed value  $\delta=0.1$:  (a) The function ${\mathcal Q_2}(\tau, \nu_0) $
		vs. $\tau$ and $\nu_0$. (b) The function $|s(\tau,\nu_0)| = \e^{ -(1/\nu_0){\mathcal Q_2}(\tau,\nu_0)}$ vs. $\tau$ and $\nu_0$.
		\label{Fig_5}}
\end{figure}

\begin{figure}
\begin{center}
\scalebox{0.19}{\includegraphics{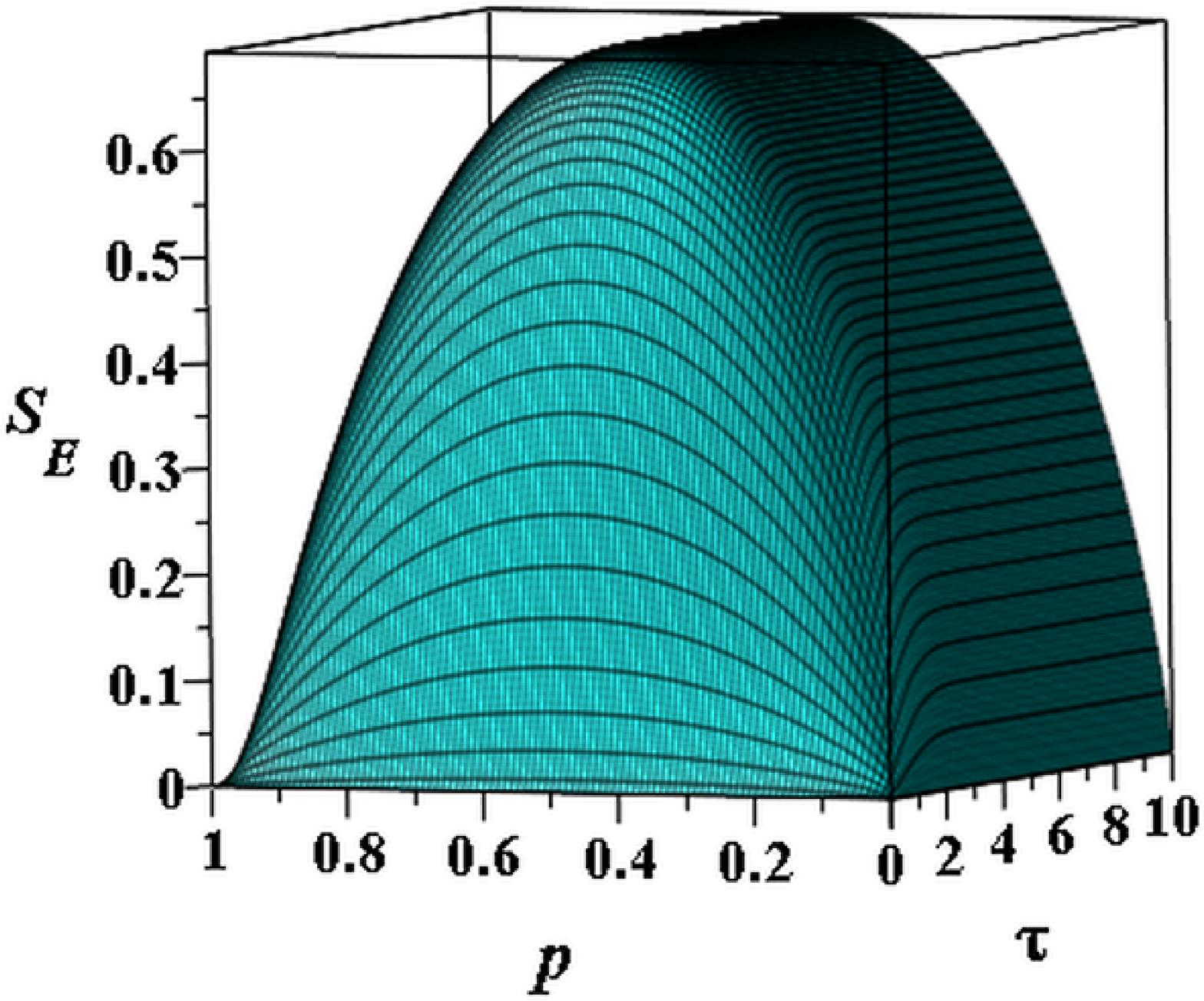}}	
(a)
\scalebox{0.19}{\includegraphics{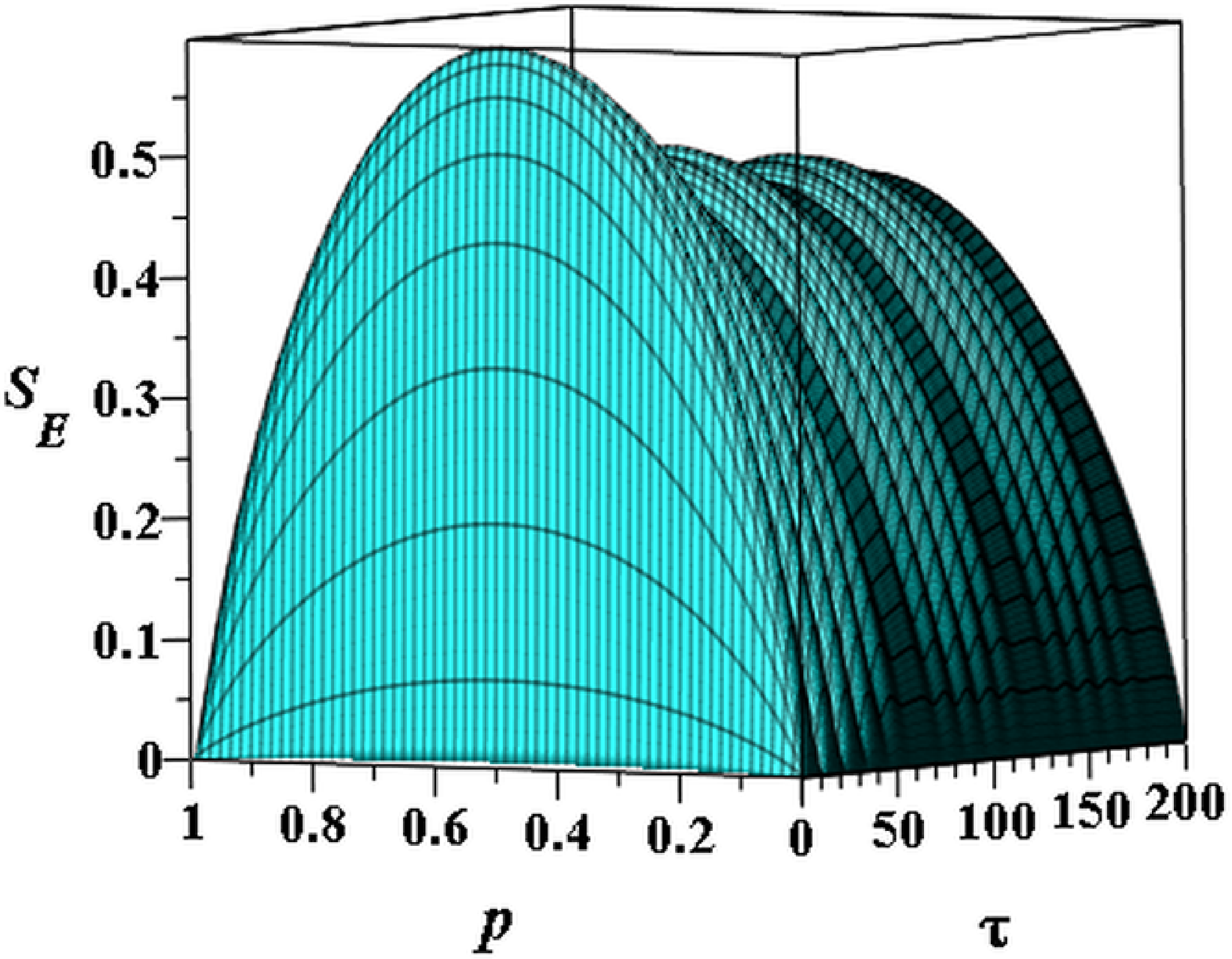}}	
(b) \\
\scalebox{0.19}{\includegraphics{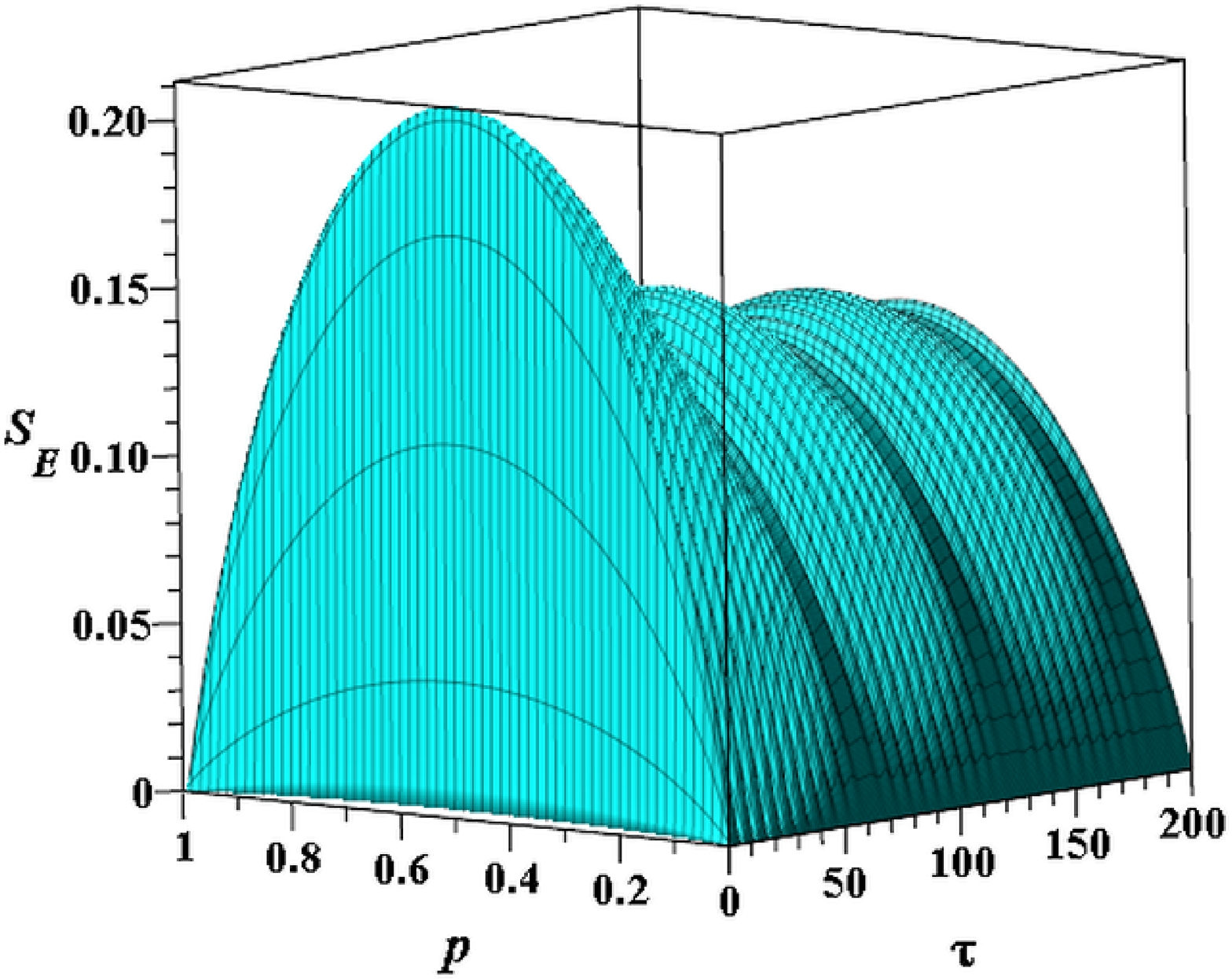}}	
(c)
\scalebox{0.19}{\includegraphics{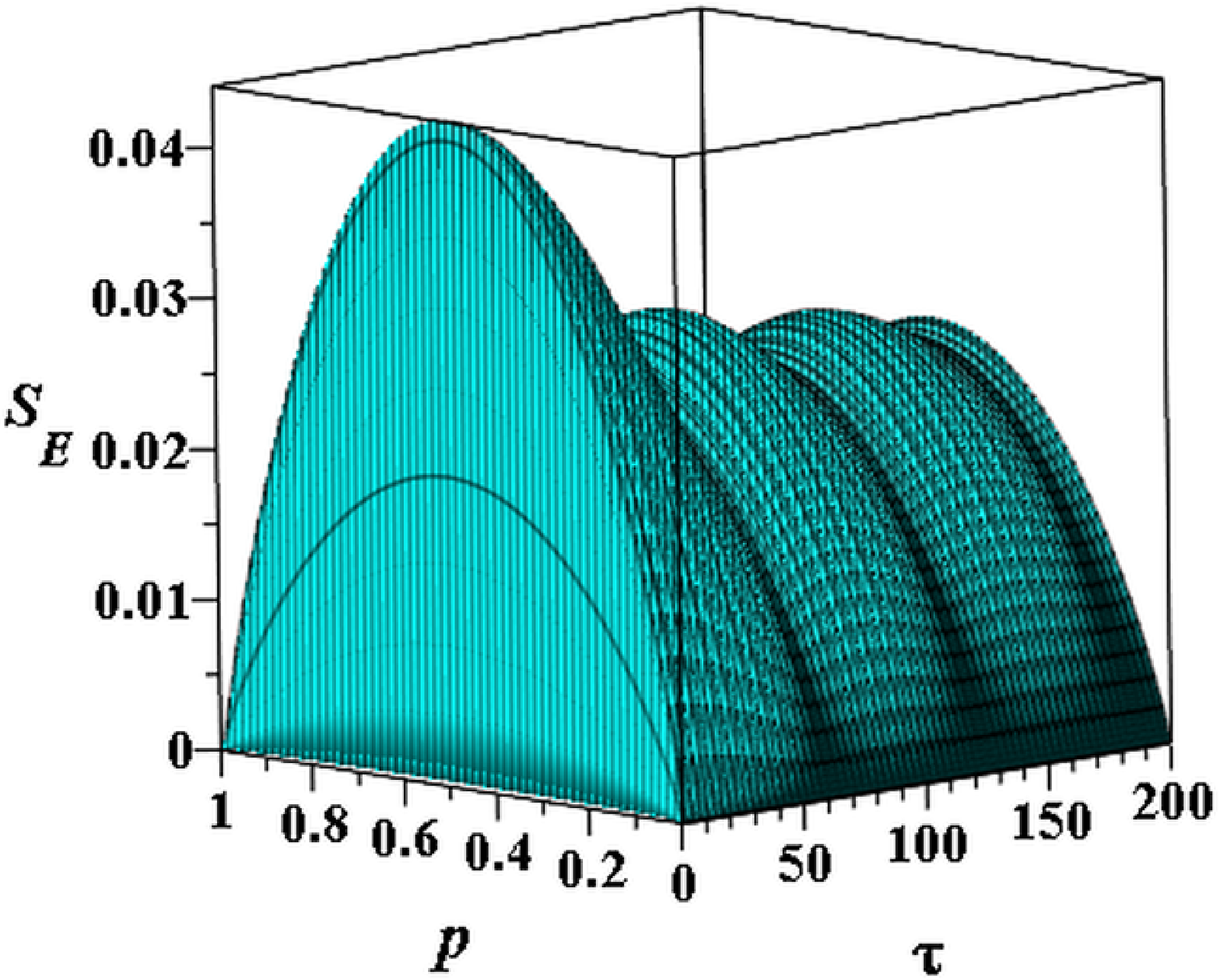}}	
(d)
\end{center}
\caption{(Color online) $X$-interaction. The entanglement entropy  $S_E$ as a 
function of $p$  and $\tau$: $ \delta=0.1$; (a) $\nu_0 =0.1$, (b) $\nu_0 =0.5$,  (c) $\nu_0 =1$, (d) $\nu_0 =2$. 
\label{Fig_6}}
\end{figure}

Let us consider the function $h(\omega)$  given by, 
 \begin{equation}	
 \label{e1}
 h(\omega) =h_0(\Theta (\omega_0)- \Theta (\omega_1)),
 \end{equation}
where $\Theta (x)$ is the Heaviside step function. We define $\omega_h=2\lambda^2\pi h_0$, and introduce the following dimensionless variables: $\nu_0 =\omega_0/\omega_h$, $\nu_1 =\omega_1/\omega_h$ and $\tau = \omega_h t$. In the new variables we have  $ |s(\tau)| = \e^{-(1/\nu_0)\mathcal Q_2(\tau )}$, where
\begin{align*}
\mathcal Q_2(\tau) =\frac{2}{\pi}\int_{\nu_0}^{\nu_1} \frac{(1-\cos(\tau z))}{z^2} \d z
= \frac{2\tau}{\pi} \big ({\rm Si}(\nu_1\tau )-{\rm Si}(\nu_0\tau ) \big )+ \frac{4}{\pi }\Big(\frac{\sin^2(\nu_0\tau/2)}{\nu_0} - \frac{\sin^2(\nu_1\tau/2)}{\nu_1} \Big).
\end{align*}
We use the asymptotic formula  ${\rm Si} (z) \sim \frac{\pi}{2} - \frac{\cos z}{z} - \frac{\sin z}{z^2}, \quad z \gg 1$, to obtain that for $\nu_0\tau,\nu_1\tau \gg 1$,
\begin{align}
	{\mathcal Q_2}(\tau) \sim \frac{2}{\pi}\Big ( \frac{\nu_1- \nu_0}{\nu_0 \nu_1  } + \frac{1}{\tau} \Big( \frac{\sin(\nu_0\tau)}{\nu^2_0} -  \frac{\sin(\nu_1\tau)}{\nu^2_1}\Big)\Big ).
\end{align}
For fixed  $ \delta =\nu_1 - \nu_0$, we study numerically the behavior of the functions $|s(\tau,\nu_0)| = \e^{-(1/\nu_0){\mathcal Q_2}(\tau,\nu_0) }$ and ${\mathcal Q_2}(\tau, \nu_0) $, where 
\begin{align}
	{\mathcal Q_2}(\tau, \nu_0) \sim \frac{2}{\pi}\Big ( \frac{\delta}{\nu_0 (\nu_0 +\delta  )} + \frac{1}{\tau} \Big( \frac{\sin(\nu_0\tau)}{\nu^2_0} -  \frac{\sin((\nu_0 +\delta )\tau)}{(\nu_0 +\delta )^2}\Big)\Big ).
\end{align}
Fig. \ref{Fig_5} shows the result for $ \delta=0.1$. The entanglement entropy as a function of $\tau$ and $p$ is depicted in Fig. \ref{Fig_6}, for a fixed value of the frequency domain, $\delta=\nu_1-\nu_0$, and for increasing value of $\nu_0$.   These results show hat the small frequencies, near $\omega=0$ give the main contribution to the EE in the X-interaction model. 

\subsection{$\mathbf D$-interaction}

\begin{figure}[h]
  \begin{center}
  \scalebox{0.2}{\includegraphics{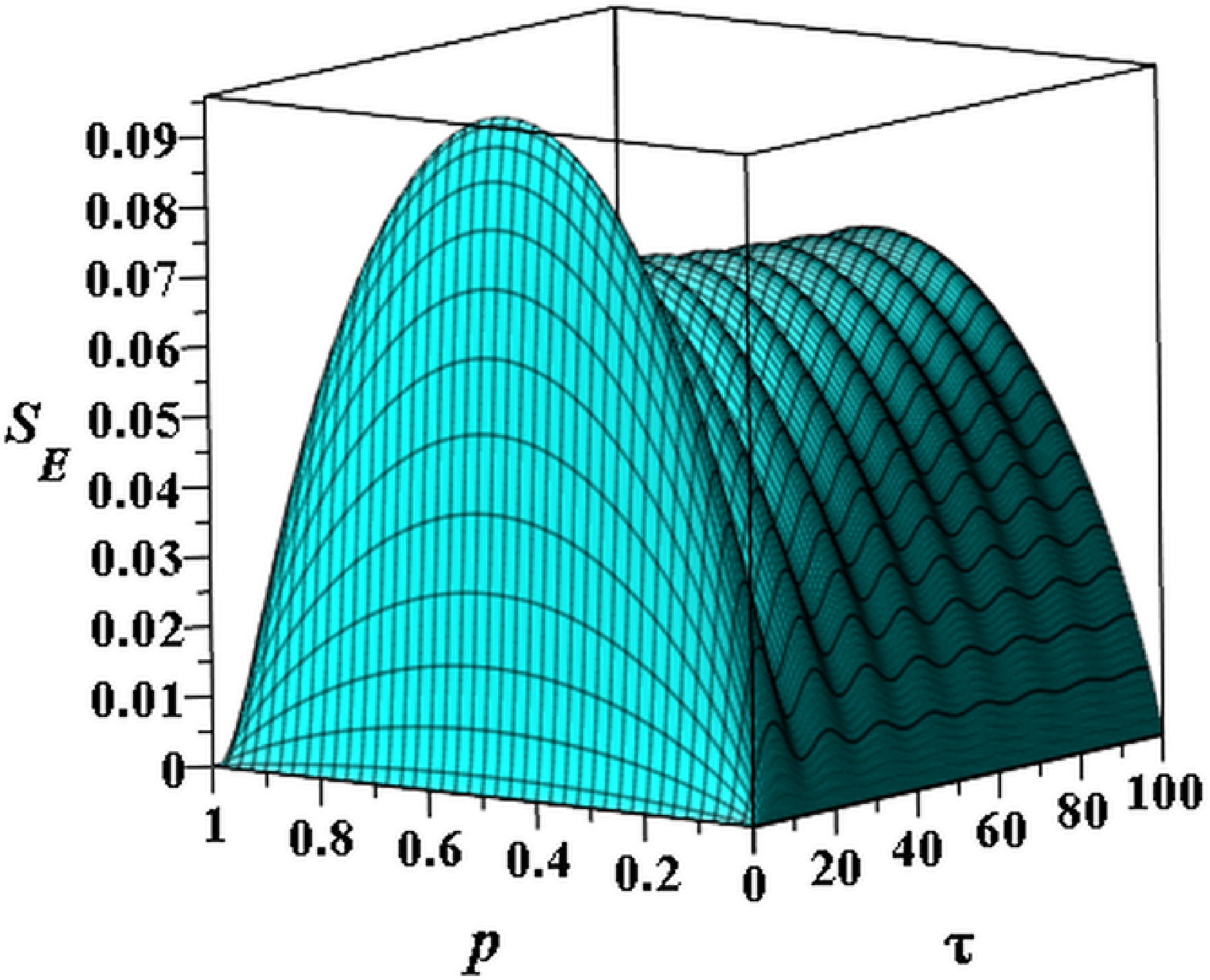}}	
  (a)
  \scalebox{0.2}{\includegraphics{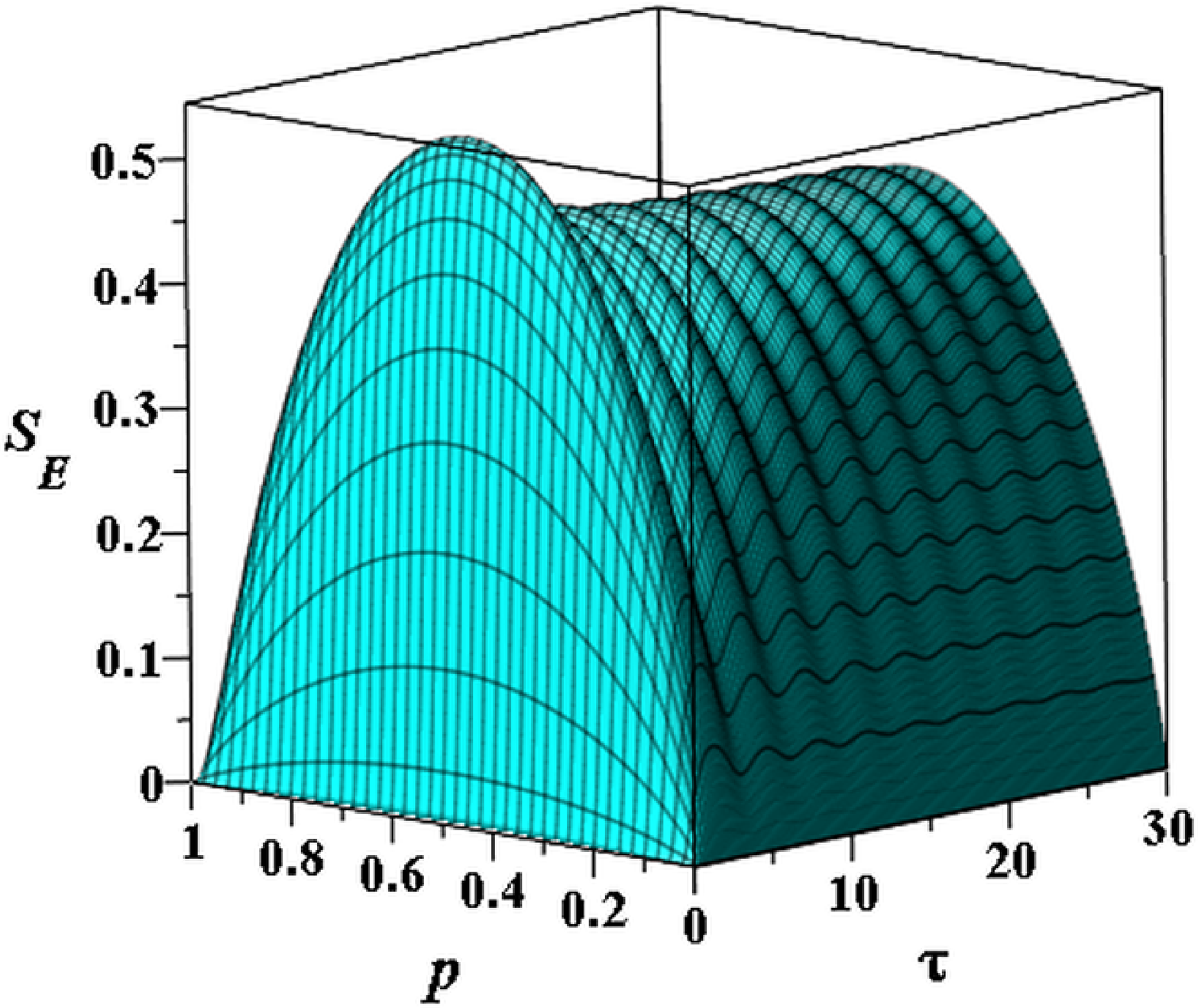}}	
  (b) \\
  \scalebox{0.2}{\includegraphics{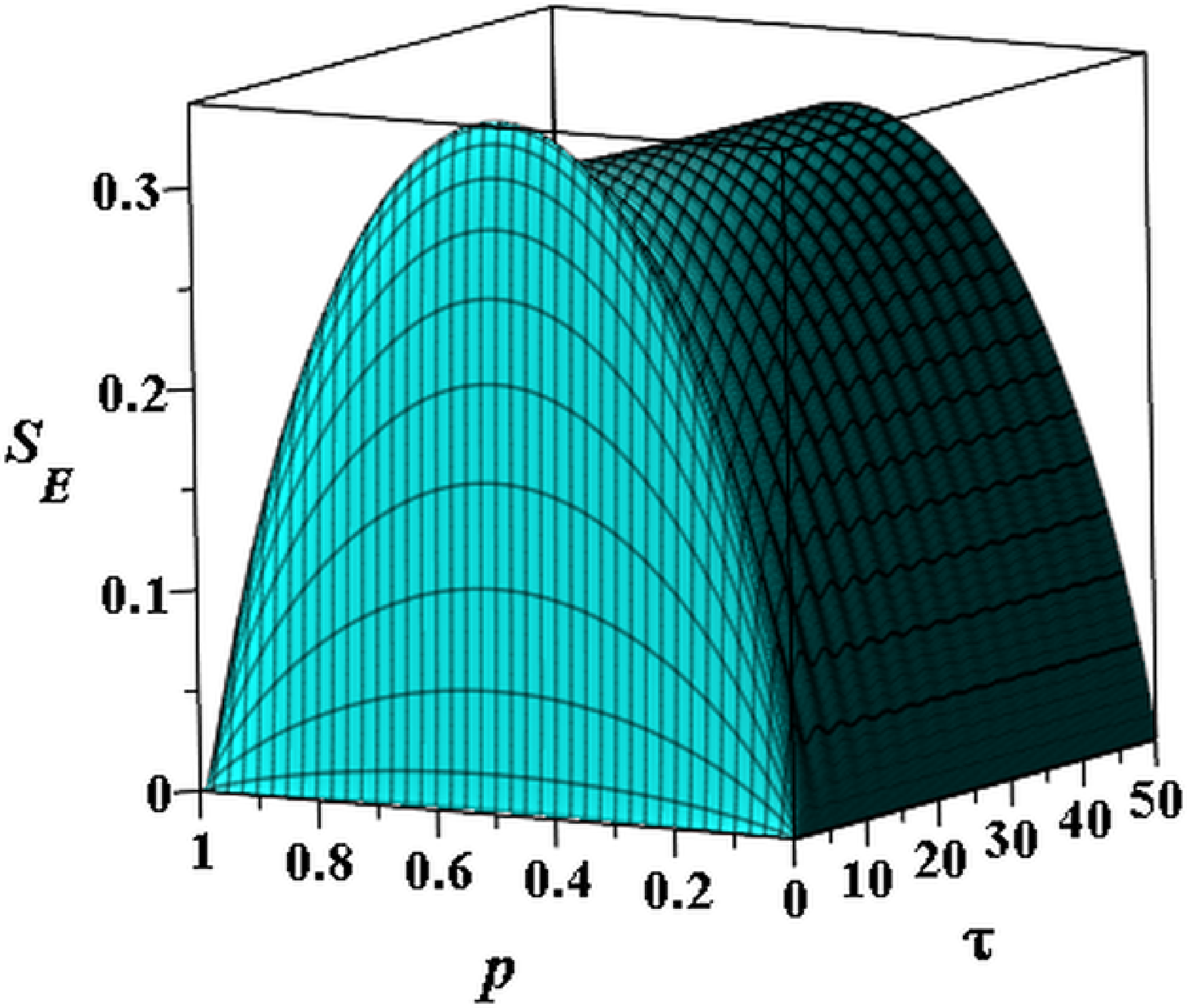}}	
  (c)
  \scalebox{0.2}{\includegraphics{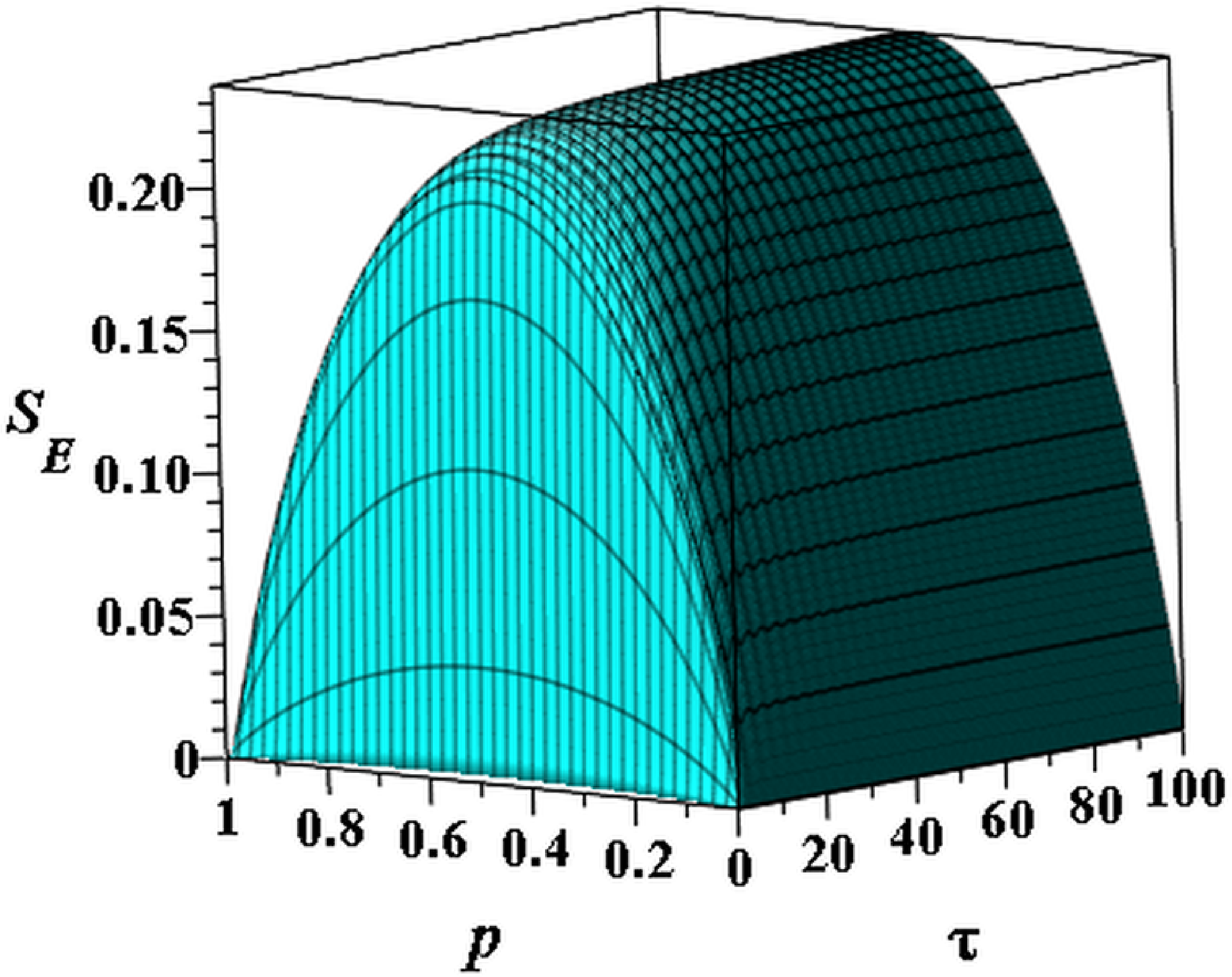}}	
  (d)
  \end{center}
 \caption{(Color online)$D$-interaction. The entanglement entropy,  $S_E$, as a  function of $p$  and $\tau$ $(\varepsilon=0.1)$. (a) $k = 1/2$, $\mu = 
 1/2$; (b) $k =1$, $\mu = 2$; 
 (c) $k = 1$, $\mu = 2$; (d) $k =2$, $\mu = 2$. 
  \label{Fig_7}}
  \end{figure}

For the $D$-interaction model, the function, $s(t)$, given by Eq. (\ref{thesj}), is written here as,
\begin{equation}
\label{SD}
s(t) = \exp \Big({-\int^\infty_0 h_D(\omega) \, [1-\e^{2\i t\lambda g(\omega)}]\, d\omega}\Big), 
	\end{equation}
For illustrative purposes, in our numerical simulations we choose $ h_D(\omega) =h_0(\Theta (0)- \Theta (\omega_c))$ and $g(\omega) =g_0 \omega^k$, where $k>0$.  Performing the integration in (\ref{SD}), we obtain,
\begin{align}
	|s(\tau)| = \exp \Big\{-\varepsilon\mu^k\Big[1-  \tfrac{1}{k(\mu 
	\tau)^{1/k}}\Re\big( \e^{-i\pi/2k}\ \gamma (1/k, i \mu
	\tau)\, \big)\Big]\Big\}, 
	\label{SD1}
\end{align}
where $\mu = 2\lambda g_0 h_0 \omega^k_c$, $\varepsilon = 
h_0 2\lambda g_0$, $\tau =h_0 t$ is the dimensionless time and 
$\gamma(x,\alpha)=\int_0^\alpha e^{-z} z^{x-1}\d z$ denotes the incomplete gamma function. Writing  $\gamma (x,\alpha) = \Gamma (x)-  \Gamma (x,\alpha)  $  and using $\Gamma(a,z) \sim z^{a-1}e^{-z}$, we obtain
 \begin{align}
 	|s(t)| \sim \exp \Big[-\varepsilon\mu^k\Big(1+ 
 	\frac{\Gamma(1/k)\cos\big(  {\pi}/{2k}\big)}{k(\mu \tau)^{1/k}}+  
 	\frac{\cos( \mu \tau) }{k(\mu\tau)^2}\Big)\Big], \quad (\mbox{ as  $\mu \tau 
 	\rightarrow \infty$}\ ).
 	\label{SD1}
 \end{align}
The entanglement entropy vs $\tau$ and $p$ is depicted in Fig. \ref{Fig_7}. As one can see, in this case, only partial decoherence occurs, and the function, $s(t)$, does not decay to zero. Correspondingly, the EE does not reach its potential maximum $\ln 2\approx 0.7$ for any values of parameters. 

\section{$X$-interaction, proofs}
\label{ecsect}

\subsection{Finite mode reservoirs}
\label{FinModeResSect}

The Hamiltonian $H_X$, \eqref{001}, leaves the dimer populations stationary. Namely, setting
\begin{equation}
\label{2}
\sigma_z  = P_1 - P_2 = |\varphi_1\rangle\langle\varphi_1| - |\varphi_2\rangle\langle\varphi_2|,
\end{equation}
where $P_1$ is the projection onto the `donor level' (state $\varphi_1$ with eigenvalue $+1$ of $\sigma_z$) and $P_2$ is the projection onto the `acceptor level' $\varphi_2$, the commutators $[H,P_1] = [H,P_2] = 0$ vanish. We take an {\em initial state} of the combined system which is pure and completely disentangled, of the form
\begin{equation}
\label{4}
\Psi(0) = \psi_{\rm S} \otimes \psi_1\otimes\psi_2 \otimes\cdots\otimes \psi_N.
\end{equation}
Here, 
\begin{equation}
\label{5}
\psi_{\rm S} = a \varphi_1 + b\varphi_2
\end{equation}
is an arbitrary pure  initial state of the dimer, with $|a|^2+|b|^2=1$ and $\psi_j$ is a pure initial state of the $j$th oscillator. Below, we specialize to the coherent state reservoir initial state \eqref{22}. We set
\begin{equation}
\label{5.1}
p=|a|^2,\qquad 1-p= |b|^2.
\end{equation}
The diagonal entries of the density matrix $|\psi_{\rm S}\rangle\langle\psi_{\rm S}|$ are $p$ and $1-p$.

\subsubsection{Reduced states}

The state of the total system at time $t$ is
\begin{equation}
\Psi(t) = \e^{-\i tH}\Psi(0)= a \, \e^{-\i\Omega t/2}P_1 \otimes_{j=1}^N U_j^+(t)\psi_j + b\ \e^{\i\Omega t/2}P_2\otimes_{j=1}^N U_j^-(t)\psi_j,
\label{6}
\end{equation}
where 
\begin{equation}
\label{7}
U_j^{\pm}(t) = \e^{-\i t\big[ \omega_j (a^\dagger_ja_j+1/2) \pm\lambda g_j(a^\dagger_j+a_j)\big]}.
\end{equation}
By a {\em polaron transformation}, we can diagonalize $U_j^\pm(t)$ as follows,
\begin{eqnarray}
D_j(\lambda g_j/\omega_j) \, U_j^+(t) \, D_j(\lambda g_j/\omega_j)^* &=& S_j(t)\equiv \e^{-\i t(\omega_j a^\dagger_ja_j - \lambda^2 |g_j|^2/\omega_j)}\nonumber\\
D_j(\lambda g_j/\omega_j)^* \, U_j^-(t) \, D_j(\lambda g_j/\omega_j) &=& S_j(t)
\label{8}
\end{eqnarray}
where $D_j(z)$, $z\in\mathbb C$, are the (coherent state) {\em displacement operators}
\begin{equation}
\label{9}
D_j(z) = \e^{ z a^\dagger_j - \bar za_j}.
\end{equation}
(Note that the second line in \eqref{8} is obtained from the first one by switching $\lambda\rightarrow-\lambda$.) 


\begin{lem}[Reduced dimer state]
	\label{lem1}
The reduced dimer density matrix has constant diagonals and (with $[\rho_{\rm S}(0)]_{12} = a\bar b$), 
	\begin{equation}
	\label{15.1}
	[\rho_{\rm S}(t)]_{12} = \e^{-\i\Omega t}\,  [\rho_{\rm S}(0)]_{12}\,\prod_{j=1}^N \scalprod{\psi_j}{D_j\big(2\lambda g_j(1-\e^{\i\omega_j t})/\omega_j\big)\psi_j}.
	\end{equation}
\end{lem}
{\em Proof.\ }
The reduced density matrix of the dimer is obtained immediately from \eqref{6},
\begin{eqnarray}
\rho_{\rm S}(t) &=& {\rm Tr}_{j=1,\ldots,N} |\Psi(t)\rangle\langle\Psi(t)| \nonumber\\
&=& |a|^2 P_1  +|b|^2 P_2+ a\bar b \,\e^{-\i\Omega t}\, |\varphi_1\rangle\langle\varphi_2| \ \prod_{j=1}^N \scalprod{\psi_j}{(U^-_j(t))^* U^+_j(t)\psi_j} + {\rm h.c.}\qquad 
\label{10}
\end{eqnarray}
Using \eqref{8} and simply writing $D_j$ for $D_j(\lambda g_j/\omega_j)$ and $S_j$ for $S_j(t)$,
\begin{equation}
\label{11}
\scalprod{\psi_j}{(U^-_j(t))^* U^+_j(t)\psi_j} =\scalprod{\psi_j}{D_j S_j^* (D_j^*)^2 S_j D_j\, \psi_j}.
\end{equation}
The displacement operators satisfy 
\begin{equation}
\label{12}
D_j(z)D_j(\zeta) = \e^{2\i{\rm Im} z\bar \zeta}D_j(z+\zeta)
\end{equation}
so that $(D_j^*)^2 = D_j(-2\lambda g_j/\omega_j)$. Next, $\e^{\i \omega_j t a^\dagger_ja_j} D_j(z) \e^{-\i\omega_j ta^\dagger_j a_j} = D_j(\e^{\i\omega_j t}z)$ and hence $S_j^*(D^*_j)^2S_j = D_j(-2\e^{\i\omega_j t}\lambda g_j/\omega_j)$. Then using again \eqref{12} twice we find
\begin{equation}
\label{13}
D_j S^*_j(D_j^*)^2 S_jD_j = D_j\big(2\lambda g_j(1-\e^{\i\omega_j t})/\omega_j\big).
\end{equation}
Using expression \eqref{13} in \eqref{11} gives  \eqref{15.1}. This shows Lemma \ref{lem1} \hfill $\blacksquare$

Next we obtain the reduced density matrix of the collection of operators with given indices.

\begin{lem}[Reduced oscillators state]
	\label{lem2}
	Let  $J\subseteq \mathbb R_+$ and denote by $\rho_J(t)$ the reduced density matrix of the oscillators having frequencies inside $J$, obtained from the total density matrix by tracing out all other oscillator degrees of freedom as well as those of the dimer. Then 
	\begin{eqnarray}
	\label{15}
	\rho_J(t) = p\  |\Psi^+_J(t)\rangle\langle \Psi^+_J(t)|\  +\  (1-p)  \ |\Psi^-_J(t)\rangle\langle \Psi^-_J(t)|,
	\end{eqnarray}
	where $p=[\rho_{\rm S}(0)]_{11}$ and 
	$$
	\Psi_J^{\pm}(t) = \otimes_{j\in \widetilde J}\,  U^\pm_j(t)\psi_j,\qquad \mbox{where $\quad \widetilde J =\big\{ j\in\{1,\ldots,N\}\ :\ \omega_j\in J \big\}$.}
	$$
	Depending on time $t$, the matrix $\rho_J(t)$ has rank either one or two. It has rank one exactly when 
	\begin{equation}
	\label{14}
	|\scalprod{\psi_j}{D_j(2\lambda g_j(1-\e^{\i t\omega_j t})/\omega_j)\psi_j}| = 1\qquad \forall j\in \widetilde J.
	\end{equation}
	Apart from $t=0$, condition \eqref{14} is difficult to satisfy, so generically, the rank of $\rho_J(t)$ is two.
\end{lem}

{\em Proof.\ } The reduced density matrix in question is obtained from \eqref{6},
\begin{eqnarray*}
	\rho_J(t) &=& {\rm Tr}_{{\rm S},\, j\in\{1,\ldots,N\}\backslash \widetilde J}\ |\Psi(t)\rangle\langle\Psi(t)|\nonumber\\
	&=& |a|^2 \otimes_{j\in \widetilde J} |U_j^+(t)\psi_j\rangle\langle U^+_j(t)\psi_j| + |b|^2 \otimes_{j\in \widetilde J} |U_j^-(t)\psi_j\rangle\langle U^-_j(t)\psi_j|.
\end{eqnarray*}
This shows the form of $\rho_J(t)$ as given in Lemma \ref{lem2}. Since the vectors $\Psi^\pm_J(t)$ are normalized, their span is one-dimensional exactly if $|\scalprod{\Psi_J^-(t)}{\Psi_J^+(t)}|=1$. This equation is equivalent to (c.f. \eqref{11}, \eqref{13}) $|\scalprod{\psi_j}{D_j(2\lambda g_j(1-\e^{\i t\omega_j t})/\omega_j)\psi_j}| = 1$,  $\forall j\in \widetilde J$.
\hfill $\blacksquare$

\subsubsection{Entanglement entropy}

\begin{lem}[Entanglement entropy]
	\label{lem3}
	The entanglement entropy of $\rho_{\rm S}(t)$ and $\rho_J(t)$, given in Lemmas \ref{lem1} and \ref{lem2}, respectively, are as follows.
	\begin{eqnarray}
	\label{21}
	S(\rho_J(t)) &=& -\big(\tfrac12+\tfrac12 r_J(t)\big)\ln\big(\tfrac12+\tfrac12 r_J(t)\big) - \big(\tfrac12 -\tfrac12 r_J(t)\big)\ln\big(\tfrac12-\tfrac12r_J(t)\big)\\
	S(\rho_{\rm S}(t)) &=& S(\rho_{J=\{\omega_1,\ldots,\omega_N\}}(t))
	\label{22.0}
	\end{eqnarray}
	where $p$ is the (time-independent) population probability of the first level of the dimer and 
	$$
	r_J(t) = \sqrt{1 -4p(1-p)(1-|s_J(t)|^2)}\mbox{\ \ and\ \ } |s_J(t)| = \prod_{j\in \widetilde J} \big|\scalprod{\psi_j}{D\big(2(1-\e^{\i\omega_j t})\lambda g_j/\omega_j\big)\psi_j} \big|.
	$$
\end{lem}

{\bf Remarks.\ } 

{\bf 1.} If the dimer is populated entirely on one level, i.e., $p=0$ or $p=1$, then $r_J(t)=1$  and $S(\rho_J(t))=0$ regardless of $J$ and $t$. That is, no entanglement entropy is created during the dynamics in {\em any} subsystem. This can be understood in the following way: for $\Psi_{\rm S} =\varphi_1$ the evolution of $\Psi(0)$, \eqref{4}, under the dynamics generated by $H_X$, \eqref{001}, is
$$
\Psi(t) =\e^{-\i tH}\Psi(0) = \e^{-\i \Omega t/2} \varphi_1\otimes \psi_1(t)\otimes\psi_2(t)\cdots\otimes\psi_N(t).
$$
Therefore, the state is of product form for all times, so the reduction to any subsystem is again a pure state and has zero entropy.

{\bf 2.} The initially pure dimer state has maximal entropy for $p=1/2$. Then $r_J(t)=|s_J(t)|$.  

\bigskip 

{\em Proof of Lemma \ref{lem3}.\ } The entanglement entropy $S(\rho_J(t)) = -{\rm Tr}\rho_J(t)\log\rho_J(t)$ of the state $\rho_J(t)$, \eqref{15}, is expressed purely through the two non-zero eigenvalues of $\rho_J(t)$. (The operator $\rho_J(t)$ acts on an infinite-dimensional Hilbert space of $|\widetilde J|$ independent harmonic oscillators, however, all but two of its eigenvalues are zero.) 

To find the two eigenvalues, we write $\rho_J = p |\Psi_J^+\rangle\langle\Psi_J^+| +(1-p) |\Psi_J^-\rangle\langle\Psi_J^-|$ (not exhibiting the $t$ dependence for simplicity of notation) and decompose 
$$
\Psi_J^- = s \Psi_J^-+\eta_J,\quad \mbox{with $s_J=\scalprod{\Psi_J^+}{\Psi_J^-}$ and $\eta_J\perp\Psi_J^+$.}
$$ 
Then we express $\rho_J$ in the ordered orthonormal basis $\{\Psi_J^+, \hat\eta_J\}$, where $\hat\eta_J = \eta_J/\|\eta_J\|$. Namely, it has the form \eqref{17.1}. The nonzero eigenvalues of $\rho_J$ are then immediately found,
\begin{equation}
\label{18}
{\rm spec}(\rho_J) = \Big\{\tfrac12 \pm \sqrt{\tfrac14 -p(1-p)(1-|s_J|^2)}\Big\}\cup \{0\}.
\end{equation}
While $p$ is the time-independent population of the first level of the dimer, the overlap $s_J$ depends on time,
\begin{equation}
\label{19}
|s_J| = \prod_{j\in \widetilde J} \big|\scalprod{\psi_j}{D\big(2(1-\e^{\i\omega_j t})\lambda g_j/\omega_j\big)\psi_j} \big|.
\end{equation}
Note that $|s_J(t)|\le 1$ and $|s_j(t)|=1 \Leftrightarrow$ $\rho_J(t)$ has rank one (c.f. Lemma \ref{lem2}). This completes the proof of Lemma \ref{lem3}. \hfill $\blacksquare$

\subsection{Coherent state reservoir}

Consider the initial reservoir state \eqref{22}.  Since $D(\xi)|\alpha\rangle = \e^{\i{\rm Im}\xi\bar\alpha}|\alpha+\xi\rangle$ we obtain $\scalprod{\alpha_j}{D(\xi_j)\alpha_j} = \e^{\i{\rm Im} \xi_j\bar\alpha_j}\scalprod{\alpha_j}{\alpha_j+\xi_j} = \e^{-2\i{\rm Im} \alpha_j\bar\xi_j}\  \e^{-\frac12|\xi_j|^2}$, where $\xi_j=2(1-\e^{\i\omega_j t})\lambda g_j/\omega_j$.
It follows that (see Lemma \ref{lem3} and \eqref{19})
\begin{equation}
\label{23''}
|s_J(t)| = \e^{-\frac12 \sum_{j\in \widetilde J} |\xi_j|^2} = \exp\Big(-2\lambda^2\sum_{j\in \widetilde J} \frac{|g_j|^2}{\omega_j^2} |1-\e^{\i\omega_j t}|^2\Big).
\end{equation}
We note that $|s_J(t)|$ is {\em independent} of the $\alpha_1,\ldots,\alpha_N$ determining the initial reservoir state! Moreover, since the entanglement entropy only depends on this quantity, it too will not depend on the $\alpha_j$.

\subsubsection{Proofs of Propositions \ref{lem4} and \ref{lem5}}
\label{contmodeslimit}

The results of Proposition \ref{lem4} follow immediately from the expressions for the discrete oscillators by using \eqref{25}. We now turn to the proof of Proposition \ref{lem5}. 

In \eqref{23},  we write $1-\cos(\omega t) = 2\sin^2(\omega t/2)$ and use the fact that 
$\lim_{t\rightarrow\infty} \frac1t \int_0^\infty h_X(\omega)\frac{\sin^2(\omega t/2)}{\omega^2} d\omega = \pi h_X(0)/4$ 
to arrive at \eqref{30}. This proves (A). The proof of (B) is obtained in the same way. To prove  (C) we use the estimate
\begin{equation}
\label{31}
\e^{-4\lambda^2 M_J\int_{\omega_0}^{\omega_1}\frac{1-\cos(\omega t)}{\omega^2} d\omega} \le |s_J(t)|\le \e^{-4\lambda^2 m_J\int_{\omega_0}^{\omega_1}\frac{1-\cos(\omega t)}{\omega^2} d\omega}.
\end{equation}
Next 
\begin{eqnarray*}
	\int_{\omega_0}^{\omega_1}\frac{1-\cos(\omega t)}{\omega^2} d\omega &\ge& \frac{1}{\omega_1^2}\int_{\omega_0}^{\omega_1}\big(1-\cos(\omega t)\big) d\omega \\
	&=&  \frac{1}{\omega_1^2}\Big( \omega_1-\omega_0 - \tfrac{\sin(\omega_1 t) - \sin(\omega_0 t)}{t}\Big) \ge \frac{1}{\omega_1^2}\Big( \omega_1-\omega_0 -2/t\Big).
\end{eqnarray*}
Combining this with \eqref{31} yields the upper bound in \eqref{28}. To get the lower bound in \eqref{28}, we use $\int_{\omega_0}^{\omega_1}\frac{1-\cos(\omega t)}{\omega^2} d\omega \le \frac{1}{\omega_0^2}(\omega_1-\omega_0+2/t)$ in \eqref{31}. This completes the proof of Proposition \ref{lem5}. \hfill $\blacksquare$

\section{$D$-interaction, proofs}
\label{dintsect}

\subsection{Dimer dynamics for $V,\mu> 0$, small} 

Theorem \ref{thm1a} below describes the change to this process under the influence of (small) energy-exchange interaction, i.e., for $V$ and $\mu$ in \eqref{0032} small. The smallness of the energy-exchange terms is measured by the  parameter
\begin{equation}
\label{66}
\eta \equiv 
\left\{
\begin{array}{cl}
\tfrac{\mu\, \sup_\omega f(\omega)}{\lambda\, \inf_\omega g(\omega)} &\mbox{if $\mu\Omega\sup_\omega f(\omega) \ge \lambda V\inf_\omega g(\omega)$}\\
\ & \\
\tfrac{V+2\mu\sup_\omega f(\omega)}{\Omega+2\lambda\inf_\omega g(\omega)} & \mbox{if  $\mu\Omega\sup_\omega f(\omega) < \lambda V\inf_\omega g(\omega)$}
\end{array}
\right.
\end{equation}
We consider here the coupling function to satisfy 
\begin{equation}
\label{emm}
	m \equiv \inf_{\omega\ge 0} g(\omega)>0.
\end{equation}
If $\mu=V=0$ then $\eta=0$.

\begin{thm}[Dynamics of the dimer]
	\label{thm1a} Let $C_\alpha = 
	\int_0^\infty h_D(\omega) f(\omega) d\omega$, where $h_D$ is given in \eqref{theh} and $f$ is the coupling function in \eqref{0032}. 
	The reduced dimer density matrix $\rho_\s(t)$ has the expansion
	\begin{eqnarray}
	\label{52}
	\big\| \rho_\s(t) - \rho_\s^0(t) - \tfrac{V}{\Omega} \e^{-\int_0^\infty h(\omega)d\omega}\rho^1_\s(t) \big\| \le C \Big( V^2/\Omega^2 + \eta   +t\eta V + t\eta\mu C_\alpha + tV^3/\Omega^2 \Big).
	\end{eqnarray} 
Here, $\rho_\s^0(t)$ is given in \eqref{23'} (with constant diagonal) and
	\begin{eqnarray}
	\label{70.1}
	\rho_\s^1(t) &=& \Big( |\!\uparrow\rangle\langle\uparrow\!|  -  |\!\downarrow\rangle\langle\downarrow\!| \Big)\ {\rm Re} \ [\rho_\s(0)]_{12} \ \big(1-\e^{-\i \Omega t}\, \big)\nonumber\\
	&& +\, |\!\uparrow\rangle\langle\downarrow\!| (p-\tfrac12)\big(1-\e^{-\i t \Omega t}\big) + \, |\!\downarrow\rangle\langle\uparrow\!| \, (p-\tfrac12)\big(1-\e^{\i t \Omega t}\,\big). 
	\nonumber
	\end{eqnarray}
In \eqref{52}, $C$ is a constant not depending on any of the parameters $V,\Omega,\eta,t,\mu$, nor on the functions $h$, $\alpha$.
\end{thm}

The remainder in \eqref{52} is {\em not uniform in time}. However, up to times $t\sim t_{\rm pd}$, when the partial decoherence is completed (c.f. \eqref{57}), the remainder is at most $O(B_1)$, where
\begin{equation}
\label{53}
B_1= V^2/\Omega^2 +\eta  +\sqrt{|\xi|}\eta V/\lambda + \sqrt{|\xi|}\eta \mu C_\alpha/\lambda.
\end{equation} 
(use $t= \sqrt{|\xi|}/\lambda$ in the right side of \eqref{52}). 
The expansion \eqref{52} is meaningful provided the remainder is smaller than the main term. Therefore, in order to resolve the  effect of the term $\propto \frac V\Omega$ on the dimer dynamics (the left side of \eqref{52}) on the time-scale $t_{\rm pd}$, we need $B_1<\!\!<\frac V\Omega$. Using the definition of $\eta$, \eqref{66} (parameter regime of the second line), we see that
\begin{equation}
\label{71}
\frac{B_1}{V/\Omega} = \frac{V}{\Omega} + \frac{1+2\frac{\mu}{V}\sup_\omega f(\omega) }{1+2\frac{\lambda}{\Omega}\inf_\omega g(\omega)} \Big[ 1+\frac{\sqrt{|\xi|}}{\lambda} ( V +\mu C_\alpha) \Big] <\!\!<1
\end{equation}
provided that
\begin{equation}
\label{72}
V<\!\!<\Omega, \qquad \sqrt{|\xi|}\, V^2<\!\!<\lambda\Omega\qquad \mbox{and}\qquad \mu\Omega\sup_\omega f(\omega) <\!\!< \lambda V\inf_\omega g(\omega).
\end{equation} 
This shows the following result.
\begin{cor}[Validity of the perturbation expansion]
	\label{cor1}
	In the parameter regime \eqref{72}, the right side of \eqref{52} is $<\!\!<V/\Omega$. This means that the perturbation expansion \eqref{52} resolves the partial decoherence process and the $O(V/\Omega)$ corrections to the dynamics, and in particular, the associated change in the dimer entanglement entropy.
\end{cor}


\bigskip
{\bf Remarks.\ } {\bf 1.} If $g'(0)\neq 0$ then 
\begin{equation}
\label{62}
\xi = \frac{h'(0) g'(0) (g^{-1})'(g(0))-h(0) g''(0) (g^{-1})'(g(0))}{[g'(0)]^2}
\end{equation}

{\bf 2.} If $\varepsilon >0$, $\delta\geq 0$ and  $g(\omega)=\varepsilon + \omega^\delta$ as $\omega\sim 0$, then we obtain
\begin{equation}
\label{60}
\xi = 
\left\{
\begin{array}{cl}
h'(0) & \mbox{if $\delta=1$}\\
2h(0) & \mbox{if $\delta=1/2$ and $h'(0)=0$}.
\end{array}
\right.
\end{equation}

Note that strictly speaking, we need $\varepsilon>0$ to have $\inf_\omega g(\omega)>0$, however, the value of $\xi$ does not depend on $\varepsilon$. This independence can be understood as follows: shifting the coupling function $g(\omega)$ in \eqref{0032} by a constant value $\varepsilon$ amounts to renormalizing the frequency $\Omega$ as $\Omega + 2\lambda\varepsilon\sum_{j=1}^Na^\dagger_ja_j$. However, the energy-conserving partial decoherence process does not depend on that frequency.

{\bf 3.} If one is only interested in the closeness of $\rho_\s(t)$ to $\rho^0_\s(t)$ then the term $O(V/\Omega)$ can be included in the remainder, and the remainder stays small for a wider parameter regime than \eqref{72}. 

{\bf 4.} We derive a more precise estimate than \eqref{52}. Namely, we can omit the remainder term $tV^3/\Omega^2$ on the right side of \eqref{52} by replacing $\rho^1_\s(t)$ in the left side of \eqref{52} by 
\begin{eqnarray}
\widetilde\rho_\s^1(t)  &=& \Big( |\!\uparrow\rangle\langle\uparrow\!|  -  |\!\downarrow\rangle\langle\downarrow\!| \Big)\ {\rm Re} \ [\rho_\s(0)]_{12} \ \big(1-\e^{-\i t\sqrt{\Omega^2+V^2}}\, \big) 	\label{70}\\
&& +\, |\!\uparrow\rangle\langle\downarrow\!| \,\Big( \big( \e^{-it \sqrt{\Omega^2+V^2}}-\e^{-i\Omega t}\big)\ [\rho_\s(0)]_{12}  +(p-\tfrac12)\big(1-\e^{-\i t \sqrt{\Omega^2+V^2}}\,\big) \Big)
\nonumber\\
&& +\, |\!\downarrow\rangle\langle\uparrow\!| \,\Big( \big( \e^{it \sqrt{\Omega^2+V^2}}-\e^{i\Omega t}\big)\ [\rho_\s(0)]_{21}  +(p-\tfrac12)\big(1-\e^{\i t \sqrt{\Omega^2+V^2}}\,\big) \Big)
\nonumber
\end{eqnarray}
Due to the bound $\e^{-it\sqrt{\Omega^2+V^2}}=\e^{-i\Omega t}(1+O(tV^2/\Omega))$, we have $\widetilde\rho_\s^1(t) = \rho_\s^1(t) +O(tV^2/\Omega)$.

\subsection{Dynamics of the oscillators for $V>0$ small, $\mu=0$}

For simplicity of the presentation, we consider in this section the Hamiltonian \eqref{0032} with $\mu=0$ only. Let $J\subseteq [0,\infty)$ be a window of continuous oscillator frequencies of interest. 

\begin{thm}[Dynamics of the oscillators]
	\label{thm02}
	Consider the system for finite $N$, with Hamiltonian \eqref{0032} and $\mu=0$. Denote by $\rho_J(t)$ the reduced density matrix of the oscillators having (discrete) frequencies inside $J\subseteq \mathbb R$. We have 
	\begin{equation}
	\label{++}
	\big\|\rho_J(t) -\rho^{00}_J(t)  - \tfrac{V}{\Omega}\rho_J^1(t)\big\|_1\le C \big( \tfrac{V^2}{\Omega}+\tfrac{tV^2}{\Omega } + \tfrac{V}{\Omega+\lambda m}\big)
	\end{equation}
	where $C$ is a constant independent of $N, J,t,V,\Omega,\lambda$. Here, $m$ is given in \eqref{emm} and $\rho^{00}_J$ and $\rho^1_J$ are as in Theorem \ref{thm01}, (B). On the left side of \eqref{++}, $\|\cdot\|_1$ denotes the trace norm of operators acting on the Hilbert space of $K$ harmonic oscillators, where $K$ is the number of discrete frequencies $\omega_j$ lying inside $J$.
\end{thm}

\subsection{Proof of Theorem \ref{thm1a} } 
The reduced dimer density matrix is 
\begin{eqnarray}
\rho_\s(t) &=& {\rm Tr}_{\rm R} \e^{-\i tH} |\psi_\s\otimes\alpha_1\cdots\otimes\alpha_N\rangle \langle \psi_\s\otimes\alpha_1\cdots\otimes\alpha_N| \e^{\i tH}\nonumber\\
&=& \sum_{n_1,\ldots,n_N \ge 0} F(n_1,\ldots,n_N)\ \e^{-\i t\bar H}|\psi_\s\rangle\langle \psi_\s| \e^{\i t\bar H},
\label{34} 
\end{eqnarray}
where
\begin{equation}
\label{39}
F(n_1,\ldots,n_N) =  \prod_{j=1}^N\e^{-|\alpha_j|^2}\frac{|\alpha_j|^{2n_j}}{(n_j)!}
\end{equation}
and
\begin{equation}
\label{35}
\bar H = \tfrac12
\begin{pmatrix}
\bar\omega & \bar V\\
\bar V & -\bar \omega
\end{pmatrix},\qquad \bar\omega = \Omega +2\lambda\sum_{j=1}^N g_j n_j,\quad \bar V = V +2\mu\sum_{j=1}^N f_j n_j.
\end{equation}
The spectral representation of $\e^{-\i t\bar H}$ is
\begin{equation}
\label{36}
\e^{-\i t\bar H} =\e^{-\i tr}|\chi_+\rangle\langle\chi_+| +\e^{\i tr}|\chi_-\rangle\langle\chi_-|,
\end{equation}
where
\begin{equation}
\label{37}
r = \tfrac12 \sqrt{\bar V^2+\bar\omega^2}
\end{equation}
and, with ${\cal N} = 2(\bar \omega^2 +\bar V^2 +\bar\omega\sqrt{\bar V^2+\bar\omega^2})$, 
\begin{equation}
\label{38}
\chi_+ ={\cal N}^{-1/2}\begin{pmatrix}
\bar\omega +\sqrt{\bar V^2+\bar\omega^2}\\
\bar V
\end{pmatrix},
\qquad 
\chi_- ={\cal N}^{-1/2}\begin{pmatrix}
-\bar V\\
\bar\omega +\sqrt{\bar V^2+\bar\omega^2}
\end{pmatrix}.
\end{equation}
In the sum in \eqref{34} we split off the term with $n_1=\ldots =n_N=0$,
\begin{equation}
\label{43}
\rho_\s(t) =  F(0)\, \e^{-\i t H^0} |\psi_\s \rangle\langle\psi_\s|\e^{\i tH^0} + \sum_{n_1,\ldots,n_N\ge 0}^{\ \ \ \ \ {}_*} F(n_1,\ldots,n_N) \ \e^{-\i t\bar H}|\psi_\s\rangle\langle \psi_\s| \e^{\i t\bar H},
\end{equation}
where $F(0)=\prod_{j=1}^N\e^{-|\alpha_j|^2}$,
\begin{equation}
\label{40}
H^0 =\tfrac12 \begin{pmatrix}
\Omega & V \\	
V & -\Omega
\end{pmatrix}
\end{equation}
and where the $*$ indicates that we omit the term in which all $n_1,\ldots,n_N$ vanish in the summation. The point of this splitting is that in the remaining sum at least one of the $n_1,\ldots,n_N$ is $\ge 1$, so that we can estimate the parameter
\begin{equation}
\label{41}
\bar\eta \equiv \frac{\bar V}{\bar \omega} 
\end{equation}
by (see also \eqref{66})
\begin{eqnarray}
\bar\eta \le\eta &\equiv&  \sup_{\nu=1,2,\ldots} \frac{V+2\mu\nu \max_jf_j}{\Omega +2\lambda  \nu \min_j g_j} \nonumber \\
& =&
\left\{
\begin{array}{cl}
\tfrac{\mu\, \max_j f_j}{\lambda\, \min_j g_j} & \mbox{if $\mu\Omega\max_j  f_j \ge \lambda V\min_j g_j$}\\
\tfrac{V+2\mu\max_j f_j}{\Omega+2\lambda\min_j g_j} & \mbox{if  $\mu\Omega\max_j f_j < \lambda V\min_j g_j$}
\end{array}.
\right.\label{42}
\end{eqnarray}
Then we use the decomposition \eqref{36} and the expansions
\begin{equation}
r = \tfrac12 \bar\omega\big(1 + O(\bar\eta^2)\big),\qquad \e^{\pm\i tr} = \e^{\pm\frac{\i}{2}t\bar\omega}\big[ 1 + O(t\bar\omega\bar\eta^2)\big] = \e^{\pm\frac{\i}{2}t\bar\omega}\big[ 1 + O(t\eta \bar V)\big]
\label{44}
\end{equation}
and
\begin{equation}
\chi_+ = |\!\uparrow\rangle + O(\eta) \equiv \begin{pmatrix}1 \\ 0\end{pmatrix} +O(\eta), \qquad	\chi_- = |\!\downarrow\rangle + O(\eta) \equiv \begin{pmatrix}0 \\ 1\end{pmatrix} +O(\eta)
\label{43.1}
\end{equation}
to estimate
\begin{equation}
\label{45}
\e^{-\i t\bar H} =\e^{-\frac\i 2 t\bar\omega}|\!\uparrow\rangle\langle\uparrow \!| + \e^{\frac\i 2  t\bar\omega}|\!\downarrow\rangle\langle\downarrow \!| +O\big(\eta + t\eta \bar V\big).
\end{equation}
It follows that
\begin{eqnarray}
\lefteqn{
	\sum_{n_1,\ldots,n_N\ge 0}^{\ \ \ \ \ {}_*} F(n_1,\ldots,n_N) \ \e^{-\i t\bar H}|\psi_\s\rangle\langle \psi_\s| \e^{\i t\bar H}}\label{46} \\
&=&
\sum_{n_1,\ldots,n_N\ge 0}^{\ \ \ \ \ {}_*} F(n_1,\ldots,n_N) \Big( 
\e^{-\frac{\i}{2}t \bar\omega} |\!\uparrow\rangle\langle \uparrow\!| + \e^{\frac{\i}{2}t \bar\omega} |\!\downarrow\rangle\langle \downarrow\!|
\Big) |\psi_\s\rangle\langle\psi_\s| 
\Big( 
\e^{\frac{\i}{2}t \bar\omega} |\!\uparrow\rangle\langle \uparrow\!| + \e^{-\frac{\i}{2}t \bar\omega} |\!\downarrow\rangle\langle \downarrow\!|
\Big) \nonumber\\
& & +Rem.,
\nonumber
\end{eqnarray}
where the remainder has the estimate (c.f. \eqref{39})
\begin{eqnarray}
|Rem.| &\le& C \Big( \eta + t\eta \sum_{n_1,\ldots,n_N\ge 0}^{\ \ \ \ \ {}_*} F(n_1,\ldots,n_N) \big[ V+ 2\mu \sum_{j=1}^N f_j n_j\big] \Big)\nonumber\\
&\le & C \Big(\eta +t\eta V +2t\eta\mu \sum_{j=1}^Nf_j |\alpha_j|^2\Big),
\label{47}
\end{eqnarray}
and where $C$ is a constant independent of $t, \eta,V,\mu,\alpha_j,N$. Finally we `remove the *' in the sum on the right side of \eqref{46} by adding (and subtracting) the summand for $n_1=\ldots=n_N=0$, and we obtain
\begin{eqnarray}
\lefteqn{
	\sum_{n_1,\ldots,n_N\ge 0}^{\ \ \ \ \ {}_*} F(n_1,\ldots,n_N) \ \e^{-\i t\bar H}|\psi_\s\rangle\langle \psi_\s| \e^{\i t\bar H}}\nonumber\\	
&=&- F(0)\, \e^{-\i t H^{00}}|\psi_\s\rangle\langle\psi_\s| \e^{\i tH^{00}} + |\!\uparrow\rangle\langle\uparrow\!| \, p + |\!\downarrow\rangle\langle\downarrow\!| (1-p)	\label{48} \\
&& +\, |\!\uparrow\rangle\langle\downarrow\!| \, a\bar b\  \e^{-\i t\Omega}\ \e^{-\sum_{j=1}^N |\alpha_j|^2(1-\e^{-2\i t\lambda g_j})}  +\, |\!\downarrow\rangle\langle\uparrow\!| \, \bar a  b\  \e^{\i t\Omega}\ \e^{-\sum_{j=1}^N |\alpha_j|^2(1-\e^{2\i t\lambda g_j})}\nonumber\\ 
&&+O\Big( \eta +t\eta V + t\eta\mu\sum_{j=1}^N f_j |\alpha_j|^2\Big).
\nonumber
\end{eqnarray}
Here we have set
\begin{equation}
\label{49}
H^{00} = \tfrac12
\begin{pmatrix}
\Omega & 0\\
0 & -\Omega
\end{pmatrix}
\end{equation}
and $\psi_\s=a|\!\uparrow\rangle +b|\!\downarrow\rangle$, $p=|a|^2$. Combining \eqref{43} and \eqref{48} yields
\begin{eqnarray}
\label{50}
\rho_\s(t) &=&  \Big( \e^{-\i t H^0} |\psi_\s \rangle\langle\psi_\s|\e^{\i tH^0} - \e^{-\i t H^{00}}|\psi_\s\rangle\langle\psi_\s| \e^{\i tH^{00}}\Big)\ \prod_{j=1}^N\e^{-|\alpha_j|^2}\nonumber\\
&& + |\!\uparrow\rangle\langle\uparrow\!| \, p + |\!\downarrow\rangle\langle\downarrow\!| (1-p)	\label{51} \\
&& +\, |\!\uparrow\rangle\langle\downarrow\!| \, a\bar b\  \e^{-\i t\Omega}\ \e^{-\sum_{j=1}^N |\alpha_j|^2(1-\e^{-2\i t\lambda g_j})} + |\!\downarrow\rangle\langle\uparrow\!| \, \bar a b\  \e^{\i t\Omega}\ \e^{-\sum_{j=1}^N |\alpha_j|^2(1-\e^{2\i t\lambda g_j})} \nonumber\\
&& +O\Big( \eta +t\eta V + t\eta\mu\sum_{j=1}^N f_j |\alpha_j|^2\Big).
\nonumber
\end{eqnarray}
Note that the main term has trace one and, for $t=0$, it reduces to the correct value $\rho_\s(0)$. We may now diagonalize $H^0$ given in \eqref{40} exactly analogously to \eqref{36}-\eqref{38}. However, we resolve the $O(V/\Omega)$-term in the expansion, namely,
\begin{equation}
\chi_+ =|\!\uparrow\rangle +\tfrac{V}{2\Omega} |\!\downarrow\rangle +O(V^2/\Omega^2), \qquad \chi_- =|\!\downarrow\rangle -\tfrac{V}{2\Omega} |\!\uparrow\rangle +O(V^2/\Omega^2).
\label{73}
\end{equation}
Using these expressions in the spectral representation $\e^{-\i tH^0} = \e^{-\i tr^0} |\chi_+\rangle\langle\chi_+| + \e^{\i tr^0} |\chi_-\rangle\langle\chi_-|$ with $r_0=\frac12\sqrt{\Omega^2+V^2}$, an easy calculation leads to
\begin{eqnarray}
\lefteqn{\e^{-\i t H^0} |\psi_\s \rangle\langle\psi_\s|\e^{\i tH^0} - \e^{-\i t H^{00}}|\psi_\s\rangle\langle\psi_\s| \e^{\i tH^{00}} }\nonumber\\
&=&  \frac{V}{\Omega} \Big( |\!\uparrow\rangle\langle\uparrow\!| - |\!\downarrow\rangle\langle\downarrow\!| \Big) \ {\rm Re}\,  [\rho_\s(0)]_{12}\big( 1-\e^{-2\i tr^0}\big)\nonumber\\
&& +\frac{V}{\Omega} |\!\uparrow\rangle\langle \downarrow\!| \Big( \big( \e^{-2\i tr^0}-\e^{-\i\Omega t}\big) [\rho_\s(0)]_{12} + (p-\tfrac12) \big(1 -\e^{-2\i tr^0}\big)\Big)\nonumber\\
&& + \frac{V}{\Omega} |\!\downarrow\rangle\langle \uparrow\!| \Big( \big( \e^{2\i tr^0}-\e^{\i\Omega t}\big) [\rho_\s(0)]_{21} + (p-\tfrac12) \big(1 -\e^{2\i tr^0}\big)\Big)\nonumber\\
&& +O(V^2/\Omega^2).\qquad 
\label{75}
\end{eqnarray}
Combining \eqref{50} with \eqref{75} and taking the continuous mode limit shows \eqref{52}. 
This finishes the proof of Theorem \ref{thm1a}. \hfill $\blacksquare$

\subsection{Proofs of Theorem \ref{thm02} and Proposition \ref{lem4}}

We examine the reduced oscillator density matrix $\rho_r$  for oscillators with indices $j=1,\ldots,r$, where  $1\le r\le N$. 

Let $H$ be the Hamiltonian \eqref{0032} with $\mu=0$ and let $|M_1,\ldots,M_r\rangle$ and $| \bar M_1,\ldots\bar M_r\rangle$ be number states of the first $r$ oscillators ($M_j,\bar M_j\in\mathbb N$). The matrix elements of $\rho_r(t)$ are
\begin{equation}
\label{01}
\scalprod{M_1,\ldots,M_r}{\rho_r(t) \bar M_1,\ldots,\bar M_r} = \sum_{n_{r+1},\ldots,n_N\in\mathbb N} F\ \scalprod{\psi_\s}{\e^{ \i tH_{\bar M}}\e^{- \i t H_ M}\psi_\s},
\end{equation}
where we recall that the initial state is \eqref{22} and where we have set
\begin{equation}
\label{02}
F = \big(\prod_{j=1}^r \scalprod{M_j}{\alpha_j}\scalprod{\alpha_j}{\bar M_j}\big)
\  \big(\prod_{j=r+1}^N| \scalprod{n_j}{\alpha_j}|^2\big) \e^{-\i t\sum_{j=1}^r \omega_j(M_j-\bar M_j)}
\end{equation}
and
\begin{equation}
\label{03}
H_M = \frac 12 \begin{pmatrix}
\Omega+\lambda T_M & V \\
V & -\Omega-\lambda T_M
\end{pmatrix},
\qquad T_M = 2\sum_{j=1}^r g_j M_j + 2\sum_{j=r+1}^N g_j n_j.
\end{equation}
Of course, $H_{\bar M}$ is  \eqref{03} with $M_j$ replace with $\bar M_j$. It is readily seen that for $V=0$, the scalar product factor in the sum of \eqref{01} is simply $p\e^{-\i \lambda t\sum_{j=1}^r g_j(M_j-\bar M_j)} + (1-p) \e^{\i \lambda t\sum_{j=1}^r g_j(M_j-\bar M_j)}$ (where $p=|\scalprod{\psi_\s}{\uparrow}|^2$). A little algebra on \eqref{01} then gives the following evolution of $\rho_r$, for $V=0$:
\begin{equation}
\label{04}
\rho_r^{(V=0)} (t) = p\  U_t^+ \rho_r(0) (U_t^+)^* +(1-p) \  U_t^-\rho_r(0) (U_t^-)^*
\end{equation}
with $\rho_r(0) = |\alpha_1,\ldots,\alpha_r\rangle\langle \alpha_1,\ldots,\alpha_r|$ and 
\begin{equation}
U_t^\pm = \e^{-\i t H^\pm},\qquad H^\pm = \sum_{j=1}^r(\omega_j\pm \lambda g_j)a^\dagger_ja_j.
\end{equation}
In particular, $\rho_r^{(V=0)}(t)$ acts non-trivially on ${\rm span}\{U_t^+ |\alpha_1,\ldots,\alpha_r\rangle, U_t^- |\alpha_1,\ldots,\alpha_r\rangle\}$ and has rank two ($t\neq 0$). By the Gram-Schmidt procedure we construct an orthonormal basis of this span and express $\rho_r^{(V=0)}(t)$ as a $2\times 2$ matrix (see \eqref{014} below).  To deal with the situation $V\neq 0$, we diagonalize the matrices $H_M$ and $H_{\bar M}$ in \eqref{01},
\begin{equation}
\label{05}
\e^{-\i t H_M} =\e^{-\i tr_M}|\chi^M_+\rangle\langle\chi^M_+| +\e^{\i tr_M}|\chi^M_-\rangle\langle\chi^M_-|.
\end{equation}
We may use the formulas \eqref{36}-\eqref{38} with suitable substitutions. In particular,
\begin{equation}
\label{06}
\chi_+^M = \begin{pmatrix}
1\\
\frac{V}{2(\Omega+\lambda T_M)}
\end{pmatrix}
+ O\big( (\tfrac{V}{\Omega})^2\big),\qquad  
\chi_-^M = \begin{pmatrix}
-\frac{V}{2(\Omega+\lambda T_M)}\\
1
\end{pmatrix}
+ O\big( (\tfrac{V}{\Omega})^2\big)\
\end{equation}
with a remainder uniform in the values of $M_1,\ldots,M_N$, and
\begin{equation}
\label{07}
\e^{\pm\i tr_M} = \e^{\pm \frac12 \i t(\Omega +\lambda T_M)}\big[ 1+O\big(t\tfrac{V^2}{\Omega +\lambda T_M}\big)\big].
\end{equation}
Our strategy is now similar to what we did in the derivation of the dimer density matrix. We split off the term in \eqref{01} in which $n_{r+1}=\cdots=n_N=0$, which we can calculate explicitly (to its main contribution). The remaining terms in the sum have at least one $n_j\ge 1$ for some $j\ge r+1$, and this will make those terms small, provided that $\min_{j\ge r+1} g_j$ is relatively large.

Using formula \eqref{05} for each exponential and expanding the eigenvectors as in \eqref{06}, it is not hard (albeit a bit tedious) to obtain the following bound,
\begin{eqnarray}
\lefteqn{\scalprod{\psi_\s}{\e^{ \i tH_{\bar M}}\e^{- \i t H_M}\psi_\s}|_{n_{r+1}=\cdots=n_N=0}}  \nonumber\\
&=&\e^{-\frac12 \i t\lambda(W_M-W_{\bar M})}\Big( p +\tfrac12\frac{V}{\Omega+\lambda W_{\bar M}} [\rho_\s(0)]_{12} + \tfrac12\frac{V}{\Omega+\lambda W_M} [\rho_\s(0)]_{21}\Big)\nonumber\\
&&+ \e^{\frac12 \i t\lambda(W_M-W_{\bar M})}\Big( 1-p -\tfrac12\frac{V}{\Omega+\lambda W_{\bar M}} [\rho_\s(0)]_{21} - \tfrac12\frac{V}{\Omega+\lambda W_M} [\rho_\s(0)]_{12}\Big) \nonumber \\
&&- \e^{\i\Omega t} \e^{\frac12 \i t\lambda(W_M+W_{\bar M})} \frac V2 \Big(\frac{1}{\Omega+\lambda W_M}-\frac{1}{\Omega+\lambda W_{\bar M}}\Big) \, [\rho_\s(0)]_{21}\nonumber \\
&&+ \e^{-\i\Omega t} \e^{-\frac12 \i t\lambda(W_M+W_{\bar M})} \frac V2 \Big(\frac{1}{\Omega+\lambda W_M}-\frac{1}{\Omega+\lambda W_{\bar M}}\Big) \, [\rho_\s(0)]_{12}\nonumber \\
&&+O\Big( \tfrac{V^2}{\Omega} +tV^2\big(\tfrac{1}{\Omega+\lambda W_M} +\tfrac{1}{\Omega+\lambda W_{\bar M}}\big)\Big).
\label{08}
\end{eqnarray}
Here, we have set $W_M = 2\sum_{j=1}^r g_j M_j$. Noticing that 
\begin{equation}
\label{09}
\e^{\frac12 \i t\lambda W_M} |M_1,\ldots,M_r\rangle = \e^{\frac12 \i t\lambda\hat W} |M_1,\ldots,M_r\rangle,\qquad \mbox{where\qquad  $\hat W = 2\sum_{j=1}^r g_j a^\dagger_ja_j$},
\end{equation}
we see that the term with $n_{r+1}=\cdots=n_N=0$ on the right side of \eqref{01} is the $\langle M_1,\ldots,M_r| \cdot |\bar M_1,\ldots,\bar M_r\rangle$-matrix element of the selfadjoint {\em operator}
\begin{eqnarray}
\e^{-\sum_{j=r+1}^N |\alpha_j|^2}\Big\{ U_t^+\Big( p \rho_r(0) +\frac V2 \rho_r(0) \frac{ [\rho_\s(0)]_{12} }{\Omega+\lambda\hat W}+ \frac V2\frac{ [\rho_\s(0)]_{21}}{\Omega+\lambda\hat W} \rho_r(0) \Big) \, (U_t^+)^*\nonumber\\
+U_t^-   \Big( (1-p)\rho_r(0) -\frac V2 \rho_r(0) \frac{ [\rho_\s(0)]_{21}}{\Omega+\lambda \hat W} - \frac V2\frac{ [\rho_\s(0)]_{12}}{\Omega+\lambda \hat W} \rho_r(0) \Big) \, (U_t^-)^*\nonumber\\
-\e^{\i \Omega t} \frac{ V[\rho_\s(0)]_{21}}{2}  U^-_t \Big( \frac{1}{\Omega+\lambda \hat W} \rho_r(0) -\rho_r(0)\frac{1}{\Omega+\lambda \hat W} \Big) (U_t^+)^*\nonumber\\
+\e^{-\i \Omega t} \frac{ V[\rho_\s(0)]_{12}}{2}  U^+_t \Big( \frac{1}{\Omega+\lambda \hat W} \rho_r(0) -\rho_r(0)\frac{1}{\Omega+\lambda \hat W} \Big) (U_t^-)^*\nonumber\\
+O\Big( \tfrac{V^2}{\Omega} +\tfrac{tV^2}{\Omega}\Big)
\label{09}
\Big\}
\end{eqnarray}
So far, the remainder in \eqref{09} is estimated in the {\em maximum norm} for matrices. More precisely, let $\rho_{r,0}(t)$ be the operator obtained by taking only the term $n_{r+1},\ldots,n_N=0$ in \eqref{01}, and let $\mu(t)$ be the operator \eqref{09}, without the error term $O$. What we have shown so far is that 
\begin{equation}
\label{maxnorm}
\sup_{M_1,\ldots,M_r, \bar M_1,\ldots \bar M_N \in\mathbb N} \langle M_1,\ldots,M_r| \big(\rho_{r,0}(t) -\mu(t)\big) | \bar M_1,\ldots \bar M_r\rangle \le C(1+t)V^2/\Omega,
\end{equation}
where $C$ is a constant independent of time $t$ and of $r$. We show now that we actually have the stronger estimate $\| \rho_{r,0} -\mu(t)\|_1\le C(1+t)V^2/\Omega$ for the same constant $C$, where $\| A\|_1={\rm Tr} |A|$ is the {\em trace norm}, with the absolute value of an operator $A$ defined by $|A|=\sqrt{A^*A}$. To do so, we recall that the error contains terms brought about by replacing $\e^{\i tH_{\bar M}}$ and $\e^{-\i tH_M}$ in the term $\scalprod{\psi_\s}{\e^{ \i tH_{\bar M}}\e^{- \i t H_ M}\psi_\s}$ in \eqref{01} by using \eqref{06} and \eqref{07}. It is then not difficult to see that this remainder is a sum of at most 16 terms, each of the form $A |\alpha\rangle\langle\alpha| B$, for some operators $A$ and $B$ satisfying $\|A\| \|B\|\le C (1+t)V^2/\Omega$, for a $C$ independent of $t, r$. To estimate this remainder in trace norm, we note that 
$$
\big| \, A|\alpha\rangle\langle\alpha|  B\, \big| = \sqrt{ B^*|\alpha\rangle\langle\alpha| \, A^* A|\alpha\rangle\langle\alpha| \, B} = \|A |\alpha\rangle\| \sqrt{ |B^*\alpha\rangle\langle B^*\alpha|} = \| A|\alpha\rangle\|\, \|B^* |\alpha\rangle\|\, |U\rangle\langle U|,
$$
where $|U\rangle$ is the normalization of $B^*|\alpha\rangle$. Hence $|( A|\alpha\rangle\langle \alpha|B) |$ is rank-one and thus its trace, being the sum of its eigenvalues, equals $\|A|\alpha\rangle\|\, \|B^*|\alpha\rangle\|\le \|A\|\, \|B^*\|\le C(1+t)V^2/\Omega$. We conclude that 
\begin{equation}
\label{tracenorm}
\|\rho_{r,0}(t)-\mu(t)\|_1\le C(1+t)V^2/\Omega.
\end{equation} 
In other words, the remainder term in \eqref{09} is understood in the trace-norm topology.

Let
\begin{equation}
\label{014}
P_0 = |0,\ldots,0\rangle\langle 0,\ldots,0|
\end{equation}
be the orthogonal projection onto the state $|n_1=0,\ldots,n_r=0\rangle$. We have $U_t^\pm P_0=P_0$, $\hat W P_0=0$ and  $P^\perp_0 \hat W\ge  m P^\perp_0$, where $m=2\min_{j\ge 1} g_j$. (Here, $P^\perp_0=\bbbone-P_0$.) Therefore, 
$$
\frac{1}{\Omega+\lambda \hat W} =\frac1\Omega P_0 +\frac{1}{\Omega+\lambda \hat W}P_0^\perp =\frac1\Omega P_0 +O\big(\tfrac{1}{\Omega+\lambda m}\big). 
$$
Using this bound in \eqref{09}, we arrive at
\begin{eqnarray}
\eqref{09} &=& \e^{-\sum_{j=r+1}^N |\alpha_j|^2}\times\nonumber\\
&& \Big\{ U_t^+p \rho_r(0)(U_t^+)^* +\frac{V[\rho_\s(0)]_{12} }{2\Omega} U_t^+\rho_r(0) P_0 + \frac{ V[\rho_\s(0)]_{21}}{2\Omega}P_0  \rho_r(0)\, (U_t^+)^*\nonumber\\
&&+U_t^-  (1-p)\rho_r(0)(U_t^-)^* - \frac{V [\rho_\s(0)]_{21}}{2\Omega}U_t^- \rho_r(0) P_0 - \frac{ V [\rho_\s(0)]_{12}}{2\Omega} P_0 \rho_r(0) \, (U_t^-)^*\nonumber\\
&&-\e^{\i \Omega t} \frac{ V[\rho_\s(0)]_{21}}{2\Omega}   \Big(P_0 \rho_r(0)(U_t^+)^* -U^-_t \rho_r(0)P_0 \Big) \nonumber\\
&&+\e^{-\i \Omega t} \frac{ V[\rho_\s(0)]_{12}}{2\Omega}  \Big(P_0\rho_r(0) (U_t^-)^* -U^+_t \rho_r(0)P_0  \Big)\nonumber\\
&&+O\Big( \tfrac{V^2}{\Omega} +\tfrac{tV^2}{\Omega} +\tfrac{V}{\Omega+\lambda m} \Big)\Big\}.
\label{09.1}
\end{eqnarray}
The error $\tfrac{V}{\Omega+\lambda m}$ will be small compared to $V/\Omega$ for sizable $m$. Again, this error is estimated in the trace-norm topology (as in \eqref{tracenorm}). 

Now we treat the terms in the sum \eqref{01} in which at least one $n_j$ is $\ge 1$. We expand $\scalprod{\psi_\s}{\e^{ \i tH_{\bar M}}\e^{- \i t H_M}\psi_\s}$ using \eqref{05}-\eqref{07}, as we did to arrive at  \eqref{08}. We allow for errors of the order $V/(\Omega+\lambda T_M)\le V/(\Omega+\lambda m)$, where $m=2\min_{1\le j\le N}g_j$. For instance, we use (c.f. \eqref{06}) $\chi_+^M=\textstyle\begin{pmatrix} 1\\0\end{pmatrix} +O(V/(\Omega+\lambda m))$.  We then obtain, whenever at least one $n_j \ge 1$ for some $j=r+1,\ldots,N$,
\begin{eqnarray}
\scalprod{\psi_\s}{\e^{ \i tH_{\bar M}}\e^{- \i t H_M}\psi_\s} = \e^{-\frac12 \i\lambda t(W_M-W_{\bar M})} p + \e^{\frac12 \i\lambda t(W_M-W_{\bar M})}  (1-p) +O\big( \tfrac{V +tV^2}{\Omega+\lambda m}  \big).
\label{010}
\end{eqnarray}
The right side of \eqref{010} is independent of $n_j$ with $j\ge r+1$. Moreover, 
\begin{eqnarray}
\sum_{n_{r+1},\ldots,n_N\in\mathbb N}\!\!\!{}^*\quad  \prod_{j=r+1}^N \big|\scalprod{n_j}{\alpha_j}\big|^2 &=& \sum_{n_{r+1},\ldots,n_N\in\mathbb N}  \prod_{j=r+1}^N \big|\scalprod{n_j}{\alpha_j}\big|^2 -  \prod_{j=r+1}^N \big|\scalprod{0}{\alpha_j}\big|^2\nonumber\\
&=& 1-\e^{-\sum_{j=r+1}^N|\alpha_j|^2},
\label{011}
\end{eqnarray}
where the star ${}^*$ indicates that we sum over all terms except the one with all $n_j$ equal to zero. Combining \eqref{010}, \eqref{011} and the definition of $F$, \eqref{02},  we see that the contribution to the sum \eqref{01} coming from all terms where at least one $n_j\geq 1$, is the $\langle M_1,\ldots,M_N| \cdot | \bar M_1,\ldots, \bar M_N\rangle$-matrix element of the {\em operator} 
\begin{equation}
\label{012}
\big(1-\e^{-\sum_{j=r+1}^N|\alpha_j|^2}\big) \Big( U_t^+ p\rho_r(0) (U_t^+)^* + U_t^- (1-p) \rho_r(0) (U_t^-)^*  \Big) +O\big( \tfrac{V +tV^2}{\Omega+\lambda m}  \big).
\end{equation}
As above, the error is estimated in the trace-norm topology. The whole sum in \eqref{01} is thus the $\langle M_1,\ldots,M_N| \cdot | \bar M_1,\ldots,\bar M_N\rangle$-matrix element of the sum of the two operators \eqref{09.1} plus \eqref{012}. We conclude that
\begin{eqnarray}
\rho_r(t) &=& U_t^+ \rho_r(0)\Big( \tfrac12 p (U_t^+)^* +v P_0 \Big) + U_t^- \rho_r(0)\Big( \tfrac12 (1-p) (U_t^-)^* - \bar v  P_0 \Big) +{\rm h.c.}  \nonumber\\
&&+O\Big( \tfrac{V^2}{\Omega} +\tfrac{tV^2}{\Omega} +\tfrac{V}{\Omega+\lambda m} \Big),
\label{013}
\end{eqnarray}
(error estimated in trace-norm topology) where we have defined
\begin{equation}
\label{016}
v = \frac V2  \frac{1-\e^{-\i\Omega t}}{\Omega}  \e^{-\sum_{j=r+1}^N|\alpha_j|^2} [\rho_\s(0)]_{12}.
\end{equation}
This shows that, modulo the remainder in \eqref{013},
\begin{equation}
\label{015}
\rho_r(t) = |e_+\rangle\langle f_+| + |e_-\rangle\langle f_-| +{\rm h.c.},
\end{equation}
where $|f_+\rangle = (\tfrac12 pU_t^++\bar v P_0)|\alpha_1,\ldots,\alpha_r\rangle$, $|f_-\rangle = (\tfrac12(1-p) U_t^- -vP_0)|\alpha_1,\ldots,\alpha_r\rangle$ and 
\begin{equation}
|e_\pm\rangle = U_t^\pm |\alpha_1,\ldots,\alpha_r\rangle.
\label{014}
\end{equation}
Since $P_0|\alpha_1,\ldots,\alpha_r\rangle\propto|0\rangle\equiv |0,\ldots,0\rangle$ this shows that (the main part of) $\rho_r(t)$ acts nontrivially on the three-dimensional space spanned by the vectors $|e_\pm\rangle$ and  $|0\rangle\equiv|0,\ldots,0\rangle$. By the Gram-Schmidt procedure we construct an orthonormal basis of ${\rm span} \{ |e_+\rangle, |e_-\rangle,  |0\rangle\}$ and then express $\rho_r(t)$ as a $3\times 3$ matrix. The orthonormal basis is given by $\{|e_+\rangle, |\hat\eta\rangle, |\hat\chi\rangle\}$, where 
\begin{eqnarray}
|\hat\eta\rangle &=& (1-|s|^2)^{-1/2} \big(|e_-\rangle-s\, |e_+\rangle \big)\nonumber\\
|\hat\chi\rangle &=& \big( 1-2 \delta \tfrac{1-{\rm Re} s}{1-|s|^2} \big)^{-1/2}\Big( |0\rangle -\delta |e_+\rangle -\delta \tfrac{1-\bar s}{\sqrt{1-|s|^2}}|\hat\eta\rangle\Big).
\end{eqnarray}
Here we have set
\begin{equation}
s=\scalprod{e_+}{e_-} = \e^{-\sum_{j=1}^r |\alpha_j|^2 (1-\e^{2it\lambda g_j})},\qquad \delta = \e^{-\sum_{j=1}^r |\alpha_j|^2}.
\end{equation}
It is now a straightforward calculation to express $\rho_t(t)$, \eqref{015}, as a matrix which in the continuous mode limit is precisely 
\eqref{rho1}. Note that for simplicity of notation, we have derived the reduced density matrix of the oscillators with indices $j=1,\ldots,r$ in this proof; the general case resulting in the reduced density matrix of oscillators with frequencies in a window $J\subseteq \mathbb R$ is directly obtained from the expressions derived here, just as we did it in Lemma \ref{lem2}. This completes the proof of Theorem \ref{thm02}. \hfill $\blacksquare$

\bigskip
{\bf Proof of Proposition \ref{lem4}.\ } We write the $3\times 3$ matrix associated to the operator $\rho_J^0(t)+\frac V\Omega\rho_J^1(t)$ (c.f. Theorem \ref{thm02}) as $M_0+\frac V\Omega M_1$ and solve the equation $\det(M_0+\frac V\Omega M_1-z)=0$ for $z$. For $V=0$ we obtain the three values $z_0= 0,\frac12\pm\frac12 r_J$, where $r_J$ is given in \eqref{rj}. Then, for each $z_0$, we make the Ansatz $z=z_0+\frac V\Omega t +\frac{V^2}{\Omega^2} t'+\cdots$ and solve the equation $\det(M_0+\frac V\Omega M_1-z)=0$ for $t$. We obtain one value for each $z_0$. \hfill$\blacksquare$

\section*{Acknowledgment}
This work was carried out under the auspices of the National Nuclear Security
Administration of the U.S. Department of Energy at Los Alamos National Laboratory
under Contract No. DE-AC52-06NA25396. M.M.  has been supported by NSERC through a Discovery Grant and a Discovery Accelerator Supplement. M.M. is grateful for the hospitality and financial support of LANL, where part of this work was carried out. A.I.N. acknowledges support from the CONACyT, Grant No. 15349 and partial support during his visit from the Biology Division, B-11, at LANL.  G.P.B. and R.T.S. acknowledge  support from the LDRD program at LANL.

\end{document}